\documentclass{article}
\usepackage{arxiv}
\usepackage[utf8]{inputenc} 
\usepackage[T1]{fontenc}    
\usepackage{hyperref}       
\usepackage{url}            
\usepackage{booktabs}       
\usepackage{amsfonts}       
\usepackage{nicefrac}       
\usepackage{microtype}      
\usepackage{lipsum}		
\usepackage{graphicx}
\usepackage{doi}

\usepackage{color}
\usepackage{tikz}
\usetikzlibrary{quantikz}
\usepackage{amssymb} 
\usepackage{caption}
\usepackage{subcaption}
\usepackage{tabularx}
\usepackage{ragged2e}

\newcommand{\Hadamard}{\text{H}}
\newcommand{\Phase}{\text{P}}

\newcommand{\Rz}{\text{R}$_\text{z}$}
\newcommand{\CX}{\text{CX}}

\newcommand{\SX}{\text{SX}}
\newcommand{\vin}{I_{in}}
\newcommand{\vouth}{I_{h}}
\newcommand{\voutv}{I_{v}}
\newcommand{\voutm}{I_{m}}
\newcommand{\voutd}{I_{d}}
\newcommand{\voutb}{I_{both}}
\newcommand{\voutt}{I_{total}}
\newcommand{\methodone}{Std32T}
\newcommand{\methodtwo}{Std50}
\newcommand{\methodthree}{Seq50}
\newcommand{\methodfour}{Para50}
\newcommand{\methodfive}{Para50\_3pix}
\newcommand{\methodsix}{SeqPara50}
\newcommand{\img}{c}
\newcommand{\weight}{w}
\newcommand{\istate}{\Theta}
\newcommand{\wstate}{\Gamma}

\newcommand{\CnX}{C^nX}

\title{A hybrid quantum image edge detector for the NISQ era}

\author{ \href{https://orcid.org/0000-0002-9955-4809}{\includegraphics[scale=0.06]{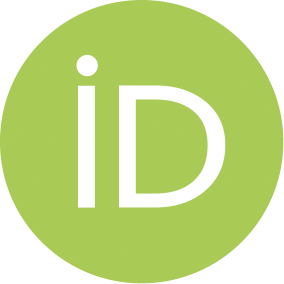}\hspace{1mm}Alexander Geng}\thanks{Corresponding author} \\
	Fraunhofer Institute for Industrial Mathematics ITWM\\ 
	Fraunhofer-Platz 1, 67663 Kaiserslautern\\            
	\texttt{alexander.geng@itwm.fraunhofer.de}\\
	\And
	\href{https://orcid.org/0000-0001-9126-3495}{\includegraphics[scale=0.06]{orcid.png}\hspace{1mm}Ali Moghiseh}\\
	Fraunhofer Institute for Industrial Mathematics ITWM\\ 
	Fraunhofer-Platz 1, 67663 Kaiserslautern\\            
	\texttt{ali.moghiseh@itwm.fraunhofer.de}\\
	\And
	\href{https://orcid.org/0000-0002-8030-069X}{\includegraphics[scale=0.06]{orcid.png}\hspace{1mm}Claudia Redenbach}\\
	University of Kaiserslautern\\ 
	Gottlieb-Daimler-Straße 47, 67663 Kaiserslautern\\            
	\texttt{redenbach@mathematik.uni-kl.de}\\
	\And
	\href{https://orcid.org/0000-0003-4903-3180}{\includegraphics[scale=0.06]{orcid.png}\hspace{1mm}Katja Schladitz} \\
	Fraunhofer Institute for Industrial Mathematics ITWM\\ 
	Fraunhofer-Platz 1, 67663 Kaiserslautern\\            
	\texttt{katja.schladitz@itwm.fraunhofer.de}\\}

\begin{document}

\maketitle

\begin{abstract}
    Edges are image locations where the gray value intensity changes suddenly. They are among the most important features to understand and segment an image. Edge detection is a standard task in digital image processing, solved for example using filtering techniques. However, the amount of data to be processed grows rapidly and pushes even supercomputers to their limits. Quantum computing promises exponentially lower memory usage in terms of the number of qubits compared to the number of classical bits. In this paper, we propose a hybrid method for quantum edge detection based on the idea of a quantum artificial neuron. Our method can be practically implemented on quantum computers, especially on those of the current noisy intermediate-scale quantum era. We compare six variants of the method to reduce the number of circuits and thus the time required for the quantum edge detection. Taking advantage of the scalability of our method, we can practically detect edges in images considerably larger than reached before.
\end{abstract}

\keywords{Quantum image processing \and quantum edge detection \and quantum artificial neurons \and IBM Quantum Experience \and real backend}

\section{Introduction}
Humans detect edges in 2D images routinely visually. In industrial applications, edge detection is used, e.g., for extracting the structure of objects, features, or regions within an image. Thereby, changes in material properties like surface defects can be detected.

In classical image processing, a standard way to highlight the edges is to compute the image gradient. It requires the pixel-wise computation of the partial gray value derivatives which is achieved by convolving the image with a filter mask assigning suitable weights to pixels in a chosen discrete neighborhood. In the filtered image, high gray values indicate a gray value change in the original, whereas low gray values indicate homogeneous neighborhoods without changes and edges. 
Various methods for edge detection have been suggested, for instance, the Prewitt, Sobel, or Laplace filters or Canny’s edge detector \cite{gonzalez2018}. The idea is to calculate pixel-wise approximations of the derivatives in vertical and horizontal direction. The filters mentioned above differ in the choice of the weights in a $3 \times 3$ filter mask. In the Canny edge detector, filtering is complemented by applying some threshold functions to suppress non-maxima and decide which components are really edges and which are rather due to noise.

Exploiting a real quantum computer, we can benefit from exponentially lower memory usage in terms of the number of qubits compared to the number of bits needed to represent an image classically. Several approaches for quantum edge detection have been proposed. However, most of them are only formulated in theory or for a quantum computer simulator \cite{mastriani2014quantum, fan2019quantum, widiyanto2019edge, ma2020demonstration, zhang2015qsobel}. When applied on real quantum computers, they  are limited by high error rates, the small number of qubits available, and low coherence times. This can lead to results too noisy to be interpretable. 

For example, in QSobel \cite{zhang2015qsobel}  -- a quantum version of the well-known classical Sobel filter -- some steps like the COPY operation or the quantum black box for calculating the gradients of all pixels can currently not be implemented. To fill these gaps is a topic of current research. The Quantum Hadamard Edge Detection algorithm was suggested as a more efficient alternative \cite{yao2017quantum}. Implementations for a state vector simulator for an $8\times 8$ pixel gray value image and for a $2\times 2$ pixel image on a real quantum computer are provided in the Qiskit textbook \cite{Qiskit-Textbook_short}. Larger image sizes are briefly discussed, too, but to our knowledge have not yet been tested in practice. 

Here, we introduce a hybrid method motivated by classical filtering and making use of Tacchino’s quantum machine learning algorithm \cite{tacchino2019artificial} and its extension to gray value images \cite{mangini2020quantum}. We use a quantum information-based cost function to compare an image patch of a test image with a binary filter mask. We perform this calculation for two filter masks highlighting vertical and horizontal edges and combine their results. With the filter mask size, we control the number of qubits and gates. For the edge detection task, we only need very few gates and by that keep the error in the current NISQ era low. In our method,  the error of each circuit is independent of the image size. This way, we can push the size of images that can be processed on the current circuit-based superconducting quantum computers of IBM \cite{ibm} to a yet unreached limit.

This paper is organized as follows. In Section~\ref{sec:method}, we roll out the strategy for solving the edge detection task by an artificial neuron and filtering in the purely classical setting. In Section~\ref{sec:preliminaries}, we explain some preliminaries needed to replace the classical artificial neuron by its quantum version in Section~\ref{section:quantum_artificial_neuron}. In Section~\ref{sec:quantum_edge_detection}, we present the idea of our quantum edge detector with 2D and 1D masks, and discuss the improvements of the version with the 1D mask in theory. We describe the experimental setup in Section~\ref{sec:qc_environment}. Experimental results are shown and discussed in Section~\ref{sec:application}. Section~\ref{sec:conclusion} concludes the paper.

\section{Method}\label{sec:method}

\subsection{Edge detection by artificial neurons}\label{sec:classical_edge_detection}
\begin{figure}[tb]
	\centering
	\begin{subfigure}[tb]{.24\linewidth}
	    \centering
        \includegraphics[width=0.4\textwidth]{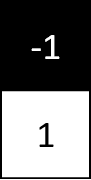}
        \caption{1D horizontal}
        \label{fig:geng_alexander:one_dim_filter_horizontal}
    \end{subfigure}
	\begin{subfigure}[tb]{.24\linewidth}
	\centering
        \includegraphics[width=0.8\textwidth]{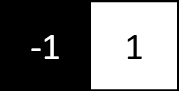}
        \caption{1D vertical}
        \label{fig:geng_alexander:one_dim_filter_vertical}
    \end{subfigure}
    \begin{subfigure}[tb]{.24\linewidth}
    \centering
        \includegraphics[width=0.8\textwidth]{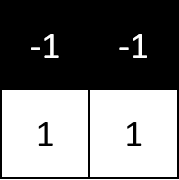}
        \caption{2D horizontal}
        \label{fig:geng_alexander:two_dim_filter_horizontal}
    \end{subfigure}
    \begin{subfigure}[tb]{.24\linewidth}
    \centering
        \includegraphics[width=0.8\textwidth]{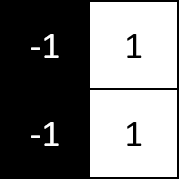}
        \caption{2D vertical}
        \label{fig:geng_alexander:two_dim_filter_vertical}
    \end{subfigure}
	\caption{One-dimensional (a,b) and two-dimensional filter masks (c,d) for horizontal and vertical direction.}
	\label{fig:geng_alexander:filter_masks}
\end{figure}

In classical image processing, we find edges of objects in an image by filtering \cite{gonzalez2018}. First- or second-order derivatives can be used to track gray value changes in the image and thereby to detect edges. Here, we opt for the first-order derivatives. We calculate digital approximations of the partial derivatives at every pixel location in the image. Let $\vin$ be a gray value input image. Then, the partial derivatives in $x$- and $y$-direction are estimated by  
\begin{equation}
    \frac{\partial \vin(x,y)}{\partial x}\approx \vin(x+1,y)-\vin(x,y)
\end{equation}
and
\begin{equation}
    \frac{\partial \vin(x,y)}{\partial y}\approx \vin(x,y+1)-\vin(x,y).
\end{equation}
This can be implemented by convolving $\vin(x,y)$ with the one-dimensional filters from Figures~\ref{fig:geng_alexander:one_dim_filter_horizontal} and \ref{fig:geng_alexander:one_dim_filter_vertical}, respectively.

Several approaches for avoiding edge effects are commonly used, e.g., zero padding, replicate padding, or mirroring, see \cite{gonzalez2018}. We use mirroring to prevent generation of artificial edges. Two-dimensional discrete derivative filters can be defined analogously, as the examples in Figures~\ref{fig:geng_alexander:two_dim_filter_horizontal} and \ref{fig:geng_alexander:two_dim_filter_vertical}. In general, there is no limit to the range of values of the weights or the size of the filter masks \cite{gonzalez2018}. However, we only cover these two filter masks for the two-dimensional case in this paper. 
Using the derivative filters for both directions, we create two output images $\vouth, \voutv$ highlighting horizontal and vertical edges, respectively. We combine them by a pixel-wise maximum:
\begin{equation} \label{equ:alexander_geng_i_both}
    \voutb=\max(\vouth,\voutv).
\end{equation}
A pixel in the image $\voutb$ has a high gray value if it belongs to an edge in horizontal or vertical direction. If neither the horizontal nor the vertical filter detects an edge, then both $\vouth$ and $\voutv$ have low values which yields a small maximum in $\voutb$. 

Finally, we segment the edge pixels by a binarization, for instance, a global gray value threshold chosen by Otsu's method \cite{gonzalez2018}. Post-processing is performed by using Fraunhofer ITWM's image processing software ToolIP \cite{toolip}. 

A special case of classical filtering is also found in the basic element of a feed-forward network, a certain kind of an artificial neural network. The goal is to learn values of weights such that a function $f$ mapping the inputs to some outputs $y$ is well approximated. Let $\img$, $\weight$ be the real valued classical input and weight vectors, respectively. The basic element of an artificial neural network is an artificial neuron --  a mathematical function, which first calculates the weighted sum of one or more inputs and then applies a non-linear activation to yield the output. It is defined by
\begin{equation} \label{equ:alexander_geng_artificial_neuron}
    y=f(\img, \weight)=\rho(\weight^T \,\img+b),
\end{equation}
where $\rho$ is the activation function and $b$ an additional so-called bias shifting the activation function for more flexibility of the network. By connecting a large number of artificial neurons in layers and by ordering layers consecutively, we can construct a feed-forward neural network. For further details, we refer to \cite{goodfellow2016deep}. 

\subsection{Quantum image processing preliminaries} \label{sec:preliminaries}
We summarize some quantum image processing preliminaries, before explaining our quantum version of the edge detector. We start at a classical image, want to find the edges in the image using a quantum computer, and finally get back a classical image in which the edges are highlighted. A key element to achieve this is encoding of the gray values of the classical image into quantum states. For encoding, unitary operations, also called gates, are applied to an initial state of the quantum computer. There are several encoding methods like basis, amplitude, or phase encoding \cite{weigold2021expanding}.

We use phase encoding to keep the number of qubits low. That means, we first transform the 8-bit gray values of an image into angles $\theta=(\theta_0, \dots,  \theta_{N-1})$ with $\theta_j\in [0,\pi]$, $j\in\{0,\dots,N-1\}$. Similar to \cite{geng2021improved}, we use the linear transformation
\begin{equation}\label{equ:theta}
    \theta_j=\img_j/255\cdot \pi, 
\end{equation}
calculated element-wise, for all $j\in\{0,\dots , N-1\}$. The transformed input vector is defined by
\begin{equation} \label{equ:geng_alexander_input_vector}
    \widetilde{\img}=(e^{i\theta_0}, e^{i\theta_1}, \dots, e^{i\theta_{N-1}}).
\end{equation}
This way, we transform the gray values to angles, encode them as phases in the quantum computer, and measure the outcome. The measurement itself is probabilistic. That means, we run the same algorithm multiple times, count the frequencies of the possible states, and derive an empirical probability distribution. The number of executions is also called number of shots. 

To compare the outcomes of the quantum computer, we use the Hellinger fidelity \cite{hellinger1909neue} derived from the Hellinger distance. Let $P,Q$ be two discrete probability distributions with probability weights $p=(p_1, \dots , p_n)$, $q=(q_1, \dots , q_n)$. Then, the Hellinger distance is defined by
\begin{equation}
    HD(P,Q)=\frac{1}{\sqrt{2}}\sqrt{\sum_{j=1}^n\left(\sqrt{p_j}-\sqrt{q_j}\right)^2},
\end{equation}
see \cite{hellinger1909neue}. The Hellinger fidelity is defined by
\begin{equation}\label{equ:hellinger_equation}
    F(P,Q)=\left(1-HD^2(P,Q)\right)^2=\left(\sum_{j=1}^n\sqrt{p_jq_j}\right)^2.
\end{equation}
It takes values in the interval $[0,1]$ with higher values for more similar distributions.

For all quantum calculations, we use the open-source software development kit Qiskit \cite{qiskit_short}.
Besides the standard gates like NOT (X), controlled-NOT (\CX), multi-controlled-NOT ($\CnX$), or Hadamard (H) gates, we also use phase shift gates (P) in this paper to encode the classical information. They have the matrix form
\begin{equation}\label{equ:phase_gate}
    P(\theta)=\left(\begin{array}{ccc}
        1 && 0\\
        0 && e^{i\theta}
    \end{array}\right),
\end{equation}
and represent a rotation around the Z-axis by an angle $\theta$ in the Bloch sphere. 
Ancilla qubits are also used, especially in conjunction with controlled operations such as the \CX\, or $\CnX$ gates. These are additional qubits which can be used for storing states, since quantum computers implement only reversible logic. At the price of more \CX\, or $\CnX$ gates, ancilla qubits can reduce the number of measurements. This is advantageous because we only need one measurement of the ancilla qubit. Thus, the structure is comparable to the classical artificial neuron with one outcome.

For the definitions and implementations of the standard gates and other basic concepts, we refer to \cite{nielsen2000quantum, Qiskit-Textbook_short}. Additionally, Table~\ref{tab:geng_alexander:basic_notions} provides some standard quantum mechanical notions from linear algebra used in this paper.
\begin{table}[tb]
    \caption{Standard quantum mechanical notions from linear algebra, similar to \cite{nielsen2000quantum}.}
    \label{tab:geng_alexander:basic_notions}
    \centering
    \begin{tabular}{cl}
        \hline\noalign{\smallskip}
        Notion & Description \\
        \noalign{\smallskip}\hline\noalign{\smallskip}
        $z^*$ & Complex conjugate of a complex number $z=a+bi$.\\
        &$z^*=(a+bi)^*=a-bi$, where $a,b\in\mathbb{R}$.\\
        $A^*$ & Complex conjugate of a matrix $A$, element-wise.\\
        $A^T$ & Transpose of a matrix $A$.\\
        $A^\dagger$ & Hermitian conjugate or adjoint of a matrix $A$, $A^\dagger=(A^T)^*$.\\
        & $\left(\begin{array}{cc} c & d\\ f & g\end{array}\right)^\dagger=\left(\begin{array}{cc} c^* & f^*\\ d^* & g^*\end{array}\right)$\\
        $\ket{\chi}$ & Encoded vector, also called ket.\\
        $\bra{\chi}$ & Dual vector, also called bra. Transposed, conjugated version of $\ket{\chi}$.\\
        &$\bra{\chi}=\left(\chi_0^*, \chi_1^*, \dots ,\chi_n^*\right)$.\\
        $\braket{\phi}{\chi}$& Inner product of $\ket{\phi}$ and $\ket{\chi}$.\\
        $\ket{\phi}\otimes \ket{\chi}$& Tensor product of $\ket{\phi}$ and $\ket{\chi}$.\\
        &$\left(\begin{array}{c} c_0 \\ c_1\end{array}\right)\otimes\left(\begin{array}{c} d_0\\ d_1\end{array}\right)=\left(c_0d_0, c_0d_1, c_1d_0, c_1d_1\right)^T$.\\
        \noalign{\smallskip}\hline
    \end{tabular}
\end{table}

\subsection{Quantum artificial neuron}\label{section:quantum_artificial_neuron}
\begin{figure}[tb]
	\centering
	\includegraphics[width=0.99\textwidth]{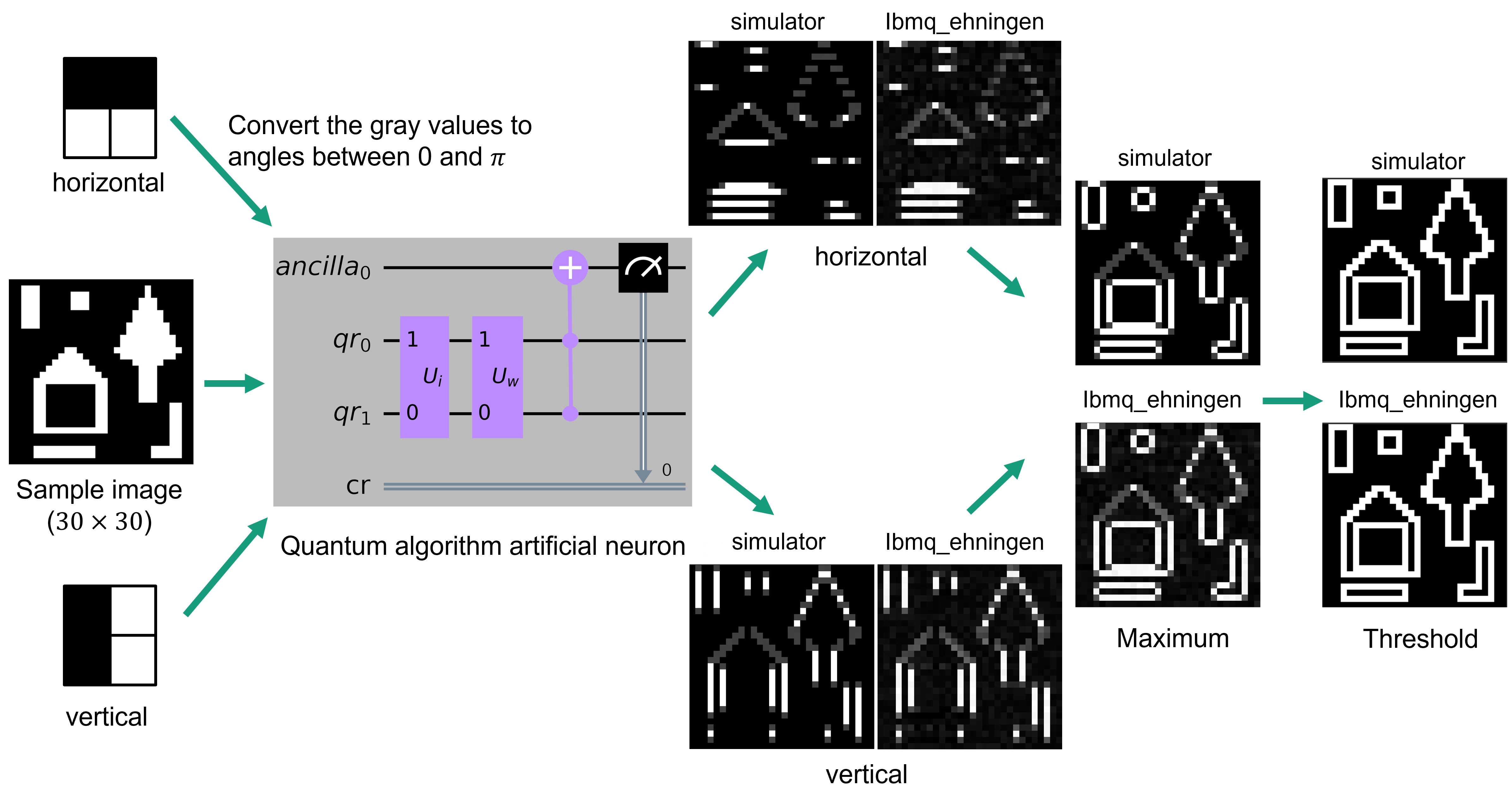}
	\caption{Scheme for edge detection in a $30 \times 30$ pixels sample image using $2\times 2$ filter masks. We use the `qasm\_simulator` and IBM's German backend `ibmq\_ehningen` (executed on November 15, 2021) with 32.000 shots, and ToolIP \cite{toolip} for post-processing.}
	\label{fig:geng_alexander:scheme}
\end{figure}

Our quantum edge detector is motivated by Tacchino's \cite{tacchino2019artificial} quantum algorithm for an artificial neuron. We will use the extension from \cite{mangini2020quantum} which also allows for treating gray value images.
Our method is sketched in Figure~\ref{fig:geng_alexander:scheme}. In the following, we explain each step, from left to right.

Let $\ket{k}$ be the $2^n$-dimensional computational basis states indexed by the decimal number $k$ corresponding to the vector of zeros and ones as binary number. We write the corresponding quantum state for the input vector $\widetilde{\img}$, see \eqref{equ:geng_alexander_input_vector}, using $n=\log_2N$ qubits
\begin{equation}\label{equ:geng_alexander_input_state}
    \ket{\istate}=\frac{1}{2^{n/2}}\sum_{k=0}^{2^n-1}\widetilde{\img_k}\ket{k}.
\end{equation}
and encode the weight vector analogously as
\begin{equation}\label{equ:geng_alexander_weight_state}
    \ket{\wstate}=\frac{1}{2^{n/2}}\sum_{k=0}^{2^n-1}\widetilde{\weight_k}\ket{k}.
\end{equation}
for weights $\gamma=(\gamma_0, \dots, \gamma_{N-1})$ with $\gamma_j\in[0,\pi]$ and corresponding vector
\begin{equation} \label{equ:geng_alexander_weight_vector}
    \widetilde{\weight}=(e^{i\gamma_0}, e^{i\gamma_1}, \dots, e^{i\gamma_{N-1}}).
\end{equation}
The inner product of the encoded input $\istate$ and weight quantum states $\wstate$ is then
\begin{equation} \label{equ:inner_product}
\begin{split}
    \braket{\wstate}{\istate} &= \frac{1}{2^n}\sum_{k,l=0}^{2^n-1}\widetilde{\img_k}\hspace{0.15cm}\widetilde{\weight_l}^*\braket{l}{k}\\
    &=\frac{1}{2^n}\widetilde{\img}^T \widetilde{\weight}^*=\frac{1}{2^n}\left(e^{i(\theta_0-\gamma_0)}+\cdots+ e^{i(\theta_{2^n-1}-\gamma_{2^n-1})}\right),
\end{split}
\end{equation}
where the second equality follows from the orthonormality of $\ket{k}$ and $\ket{l}$. 
Thus, the calculation corresponds to the scalar product of the input vector from \eqref{equ:geng_alexander_input_vector} and the conjugated weight vector from \eqref{equ:geng_alexander_weight_vector}, analogously to the classical artificial neuron. We set $b=0$ in Equation~\eqref{equ:alexander_geng_artificial_neuron} and $\rho(\cdot)= \lvert \cdot \rvert^2$ is the activation function of the quantum neuron.

To encode the inner product, unitary operations/gates have to be applied. In quantum computing, the qubits are usually initialized in well-prepared states. First, we transform this initial state into the input quantum state by the unitary operation $U_I$. The following operation $U_W$ yields the inner product of input and weight quantum state. Via a
multi-controlled-NOT ($\CnX$) gate targeting an ancilla qubit and controlled by $n$ qubits, we extract the result by measuring the ancilla qubit.

In Qiskit \cite{qiskit_short}, the state $\ket{0}$ is the initial state for all qubits. Thus, the $n$-qubit state at the beginning is $\ket{00\dots0}=\ket{0}^{\otimes n}$, where $\otimes$ stands for the tensor product. The operation $U_I$ creates the input quantum state 
\begin{equation}
    U_I\ket{0}^{\otimes n}=\ket{\istate}
\end{equation}
as given in \eqref{equ:geng_alexander_input_state}. 

It can be built in two steps. First, we apply Hadamard gates $H^{\otimes n}$ to the qubits, to create a balanced superposition state $\ket{+}^{\otimes n}$ with $\ket{+}=(\ket{0}+\ket{1})/\sqrt{2}$. 

Second, the appropriate phase has to be added to the equally weighted superposition of all the states in the $n$ qubits computational basis, in order to obtain $\ket{\istate}$. This corresponds to the diagonal unitary operation
\begin{equation}
    U(\theta)=\left(\begin{array}{cccc}
        e^{i\theta_0}   &   0           &\cdots& 0 \\
        0               &e^{i\theta_1}  &\cdots& 0 \\
        \vdots          &   \vdots      &\ddots& \vdots \\
        0               &   0           &\cdots& e^{i\theta_{2^n-1}}
    \end{array}\right).
\end{equation}
Instead of calculating the complete unitary matrix $U(\theta)$, we decompose it into 
\begin{equation}
    U(\theta)=\prod_{j=0}^{2^n-1}U(\theta_i)
\end{equation}
with $U(\theta_j)\ket{j}=e^{i\theta_j}\ket{j}$. With one $U(\theta_j)$, we apply a phase shift to one computational basis state and leave all the other states unchanged. Practically, this is realized by a combination of X-gates (to which state we want to apply a phase shift) and a multi-controlled phase shift gate $C^{n-1}P( \theta)$ as defined in \eqref{equ:phase_gate} for $n=1$. In total, we have
\begin{equation}
    U_I\ket{0}^{\otimes n}=U(\theta)H^{\otimes n}\ket{0}^{\otimes n}=\ket{\istate}.
\end{equation}
The unitary $U_W$ is encoded similarly, just conjugated, see \cite{mangini2020quantum}. Consequently, the actual prepared quantum state is 
\begin{equation}
    \ket{\phi}=(U(\gamma)H^{\otimes n})^\dagger\ket{\istate}.
\end{equation}

To extract the results, we apply X-gates $X^{\otimes n}$ to the qubits, such that the desired coefficient is the one of the state $\ket{1}^{\otimes n}$. This step completes the unitary operator 
\begin{equation}
    U_W=X^{\otimes n}H^{\otimes n}U(\gamma)^\dagger.
\end{equation}
Finally, we use an ancilla qubit as in \cite{mangini2020quantum} and map the result to it by a multi-controlled-NOT gate ($\CnX$) with $n$ control qubits
\begin{equation}
    \CnX(X^{\otimes n}\ket{\phi}\ket{0}_a)=\sum_{j=0}^{2^n-2}r_j\ket{j}\ket{0}_a+r_{2^n-1}\ket{11\cdots 1} \ket{1}_a,
\end{equation}
where $r_{2^n-1}=\braket{\wstate}{\istate}$.

\begin{figure}[tb]
	\centering
	\includegraphics[width=0.99\textwidth]{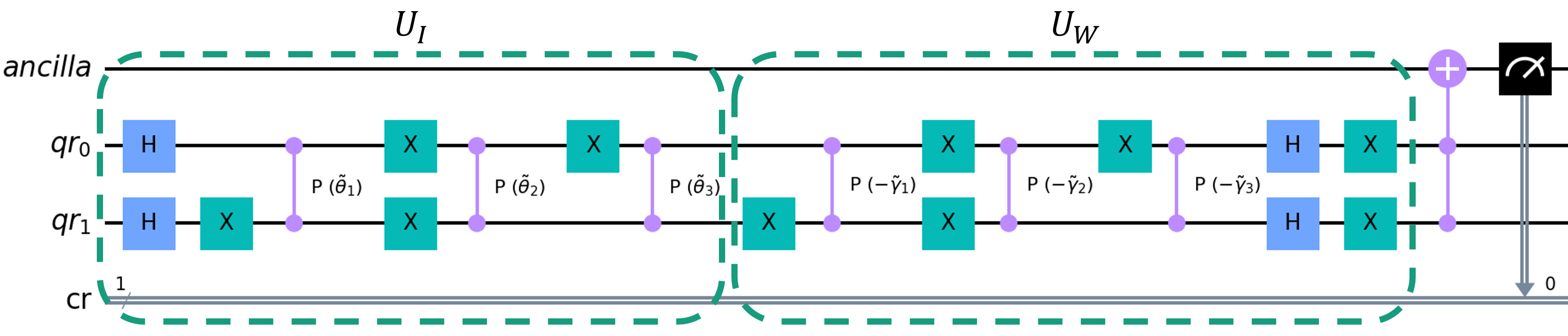}
	\caption{Quantum circuit for a $2\times 2$ input image patch (encoded in $U_I$) and $2\times 2$ filter mask (encoded in $U_W$). Two qubits plus an additional ancilla qubit are needed. The redefined pixel values for the input are encoded in angles $\tilde{\theta}_j$ and the filter mask weights in $\tilde{\gamma}_j$, for $j\in\{1,2,3\}$. It holds $\tilde{\theta}_j=\theta_j-\theta_0$ and $\tilde{\gamma}_j=\gamma_j-\gamma_0$ for $j\in\{1,2,3\}$.}
	\label{fig:geng_alexander:circuit_2x2}
\end{figure}

Probabilities in quantum mechanics are represented by the squared modulus of wave function amplitudes (Born rule \cite{born1926quantenmechanik}). This fact, combined with the global phase invariance, yields the activation function
\begin{equation} \label{equ:alexander_geng_global_phase}
    \lvert\braket{\wstate}{\istate}\rvert^2=\frac{1}{2^{2n}}\left \lvert\sum_{j=0}^{2^n-1}e^{i(\theta_j-\gamma_j)}\right\rvert^2=\frac{1}{2^{2n}}\left \lvert1+\sum_{j=1}^{2^n-1}e^{i(\tilde{\theta}_j-\tilde{\gamma}_j)}\right \rvert^2,
\end{equation}
where $\tilde{\theta}_j=\theta_j-\theta_0$ and $\tilde{\gamma}_j=\gamma_j-\gamma_0$ for $j\in\{1,\dots,2^n-1\}$. Figure~\ref{fig:geng_alexander:circuit_2x2} shows the circuit for $n=2$ qubits. 

\subsection{Quantum edge detection} \label{sec:quantum_edge_detection}
\subsubsection{Quantum edge detection with 2D mask}
In order to use the idea of the quantum artificial neuron of the previous section for quantum edge detection, we have to split the input image into $2\times 2$ patches. The vectorized version of this patch serves as input vector $\img$ and the vectorized version of the 2D mask as weight vector $\weight$. The main idea of our quantum edge detection is to replace the classical calculation of the inner product by the quantum artificial neuron. All the other classical steps from Section~\ref{sec:classical_edge_detection}, like selecting vertical and horizontal directions, combining them, and applying a threshold, remain the same and are calculated on a classical computer. 

\subsubsection{Quantum edge detection with 1D mask}
\begin{figure}[tb]
	\centering
	\includegraphics[width=\textwidth]{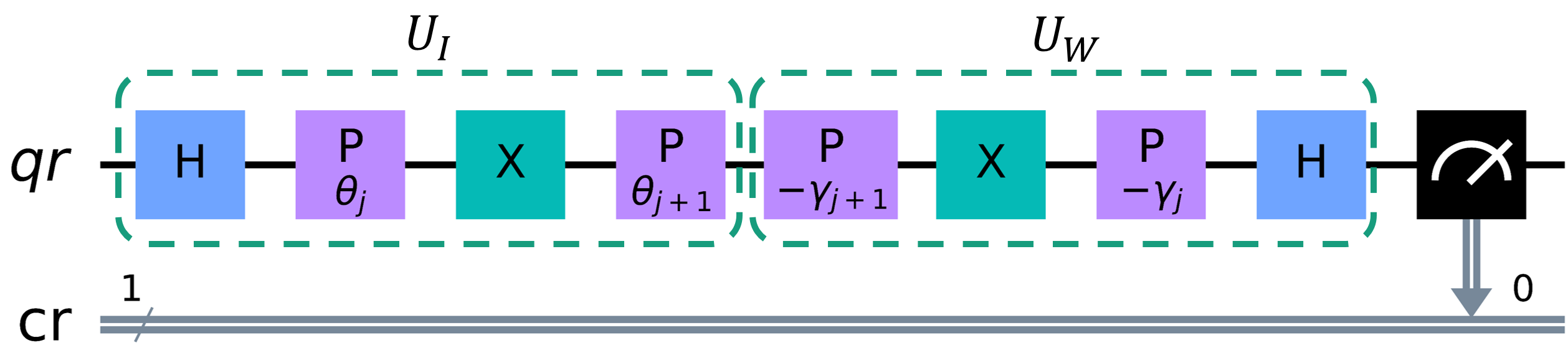}
	\caption{Quantum circuit for a two pixel input image patch (encoded in $U_I$) and two pixel filter mask (encoded in $U_W$). The redefined pixel values for the input are encoded in angles $\theta_j, \theta_{j+1}$ and the filter mask weights in $\gamma_j, \gamma_{j+1}$. By that, we get the one-dimensional filter as visualized in Figures~\ref{fig:geng_alexander:one_dim_filter_horizontal} and \ref{fig:geng_alexander:one_dim_filter_vertical}. The \Hadamard- and \Phase-gates are converted further into basis gates in the transpilation step. In total, two \SX- and three \Rz-gates are needed for this circuit.}
	\label{fig:geng_alexander:circuit_1x2}
\end{figure}

For the sake of generality, we chose an approach motivated by classical 2D filtering. However, if we are interested in edges only, then we can also use one-dimensional filter masks as in Figures~\ref{fig:geng_alexander:one_dim_filter_horizontal} and \ref{fig:geng_alexander:one_dim_filter_vertical}. This is advantageous since we only have to encode two classical pixel values into a quantum state. 
We only need one qubit for that and fewer gates compared to the two-dimensional filtering described above. That way, the algorithm is much less error-prone. A circuit for the one-dimensional case is shown in Figure~\ref{fig:geng_alexander:circuit_1x2}.

Only two Hadamard gates (\Hadamard), four Phase gates (\Phase), and two X gates are required. Analytically, we describe the circuit by
\begin{equation}\label{equ:analytical_description}
    \begin{aligned}
         U(\theta, \gamma)\ket{0}&=U_W U_I\ket{0}=HP(-\gamma_j)XP(-\gamma_{j+1})P(\theta_{j+1})XP(\theta_j)H \ket{0}\\
         &=0.5\left(e^{i\lambda_{j+1}}+e^{i\lambda_j}\right)\ket{0}+ 0.5\left(e^{i\lambda_{j+1}}-e^{i\lambda_j}\right)\ket{1},
    \end{aligned}
\end{equation}
where $\lambda_j=\theta_j-\gamma_j$ and $\lambda_{j+1}=\theta_{j+1}-\gamma_{j+1}$.
Note, that the angles for the filter mask weights in $U_W$ are the negative of the original one and the order of the gates in $U_W$ is completely opposite to $U_I$ (see \eqref{equ:inner_product} and \cite{mangini2020quantum}). At the end, we measure the qubit in state $\ket{0}$ and get the result
\begin{equation}\label{equ:analytical_result}
    \lvert\bra{0}U(\theta,\gamma)\ket{0}\rvert^2=\frac{1}{4}\lvert e^{i\lambda_{j+1}}+e^{i\lambda_j}\rvert^2,
\end{equation}
where $\theta=(\theta_j, \theta_{j+1})$ and $\gamma=(\gamma_j, \gamma_{j+1})$.

On a real backend, all unitary gates of the circuits are transformed to the basis gates. Here, this results in a decomposition into three \Rz- and two \SX-gates and a circuit depth of six. Other gates, in particular the error-prone \CX-gates are not required. For the two-dimensional case shown in Figure~\ref{fig:geng_alexander:circuit_2x2}, we need around $31$ \Rz-, $22$ \CX-, $6$ \SX-, $5$ NOT-gates, and have approximately a circuit depth of $49$ depending on the coupling map and the configuration of the chosen backend.

A drawback of the quantum edge detection with 1D mask are some missing elements in the image, as visible in Figure~\ref{fig:geng_alexander:result_binary_own_one_dim} in Section~\ref{sec:results_one_dim}. To circumvent this problem, an additional direction is required. We add input image patches for the diagonal direction 
$\vin(x,y)$, $\vin(x+1,y+1)$ of the input image. The outcome $\voutd$ is combined with those for the vertical and horizontal directions by the pixel-wise maximum yielding
\begin{equation} \label{equ:max_total}
    \voutt=\max_{m\in \{h,v,d\}}\voutm.
\end{equation}

\subsubsection{Improved Quantum edge detection with 1D mask} \label{sec:improved_1D}
Using the 1D masks, we need fewer gates, and therefore also observe less noise. Now we modify this solution in several ways to further reduce the numbers of circuits and jobs and to shorten the execution time. We compare six variants of the implementation in the following. The first one, denoted by \methodone, is the one-dimensional variant with $32.000$ shots from above. In the second one, \methodtwo, we decrease the number of shots to $50$ while the method and the circuit remain the same. 

\begin{figure}[tb]
  \begin{minipage}{.33\linewidth}
    \includegraphics[width=0.99\textwidth]{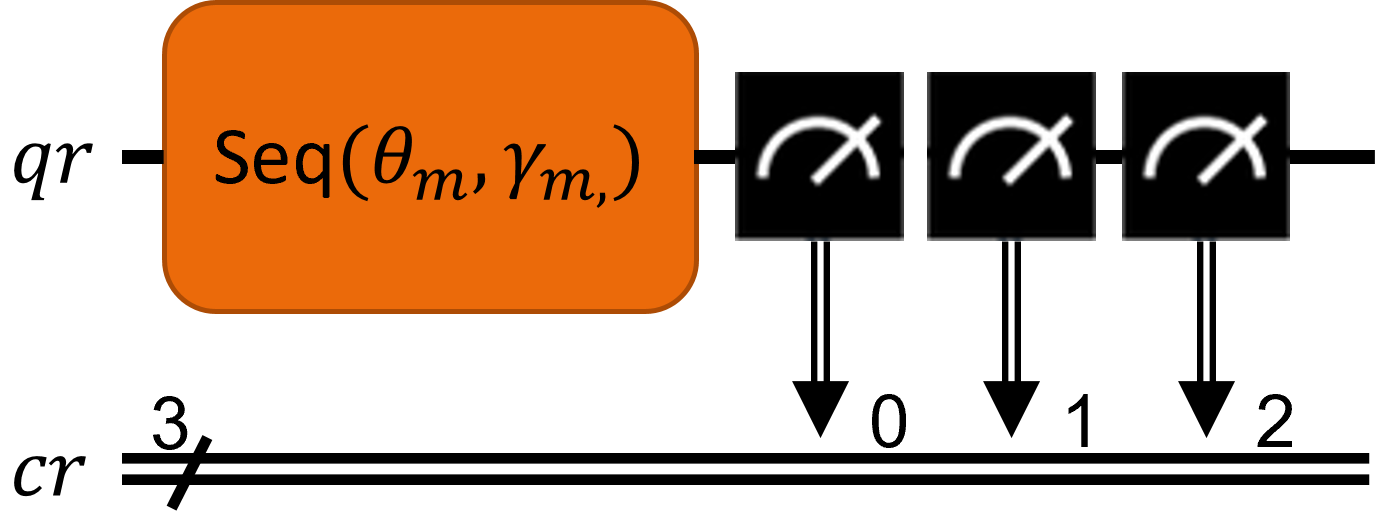}
  \end{minipage}%
  \begin{minipage}{0.07\linewidth}
    \vspace{-0.5cm}
    \begin{eqnarray*}
       \hspace{0.3cm}\equiv
    \end{eqnarray*}
  \end{minipage}%
  \begin{minipage}{.59\linewidth}
    \includegraphics[width=0.99\textwidth]{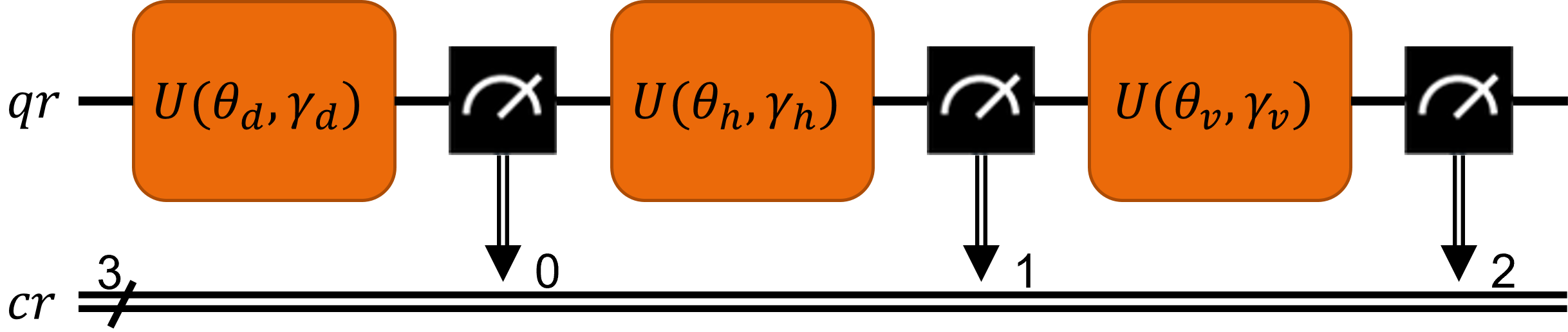}
  \end{minipage}
  \caption{Circuit scheme for \methodthree. Quantum circuit for a two pixel input image patch. The first qubit encodes the pixels in diagonal, the second those in horizontal, and the third those in vertical direction. We have $\gamma_m=(0,\pi)$, for all $m\in\{h,v,d\}$, since in all cases the first pixel of the filter mask is black and the second white (see Figures~\ref{fig:geng_alexander:one_dim_filter_horizontal} and \ref{fig:geng_alexander:one_dim_filter_vertical}). The \Hadamard- and \Phase-gates  must be converted further into basis gates. In total, six \SX- and nine \Rz-gates are needed for this circuit.}
  \label{fig:geng_alexander:circuit_mid-circuit}
\end{figure}

The remaining four variants of the one-dimensional quantum edge detector involve mid-circuit measurement, parallelism, and also 50 measurements, and are dedicated to detecting edges in larger images  using less circuits to be applicable on the current quantum computers. 
So far, we have to execute each of the three directions (horizontal, vertical, and diagonal) separately. In the third variant, \methodthree, we combine all three directions in one circuit sequentially by using mid-circuit measurements allowing qubits to be individually measured at any point in the circuit. IBM launched this feature of their backends in the beginning of $2021$ \cite{ibm}. We use it to measure the required qubit three times, once for each direction. Figure~\ref{fig:geng_alexander:circuit_mid-circuit} shows this variant.
With this improvement, we decrease the number of circuits by a factor of $3$. To retrieve the results for the three directions, we marginalize the counts from the experiment over the three indices (0 for diagonal, 1 for horizontal, and 2 for vertical). For that, we use Qiskit's utility function \textit{marginal\_counts} \cite{qiskit_short}.

\begin{figure}[tb]
    \centering
    \begin{minipage}{.38\linewidth}
    \includegraphics[width=0.99\textwidth]{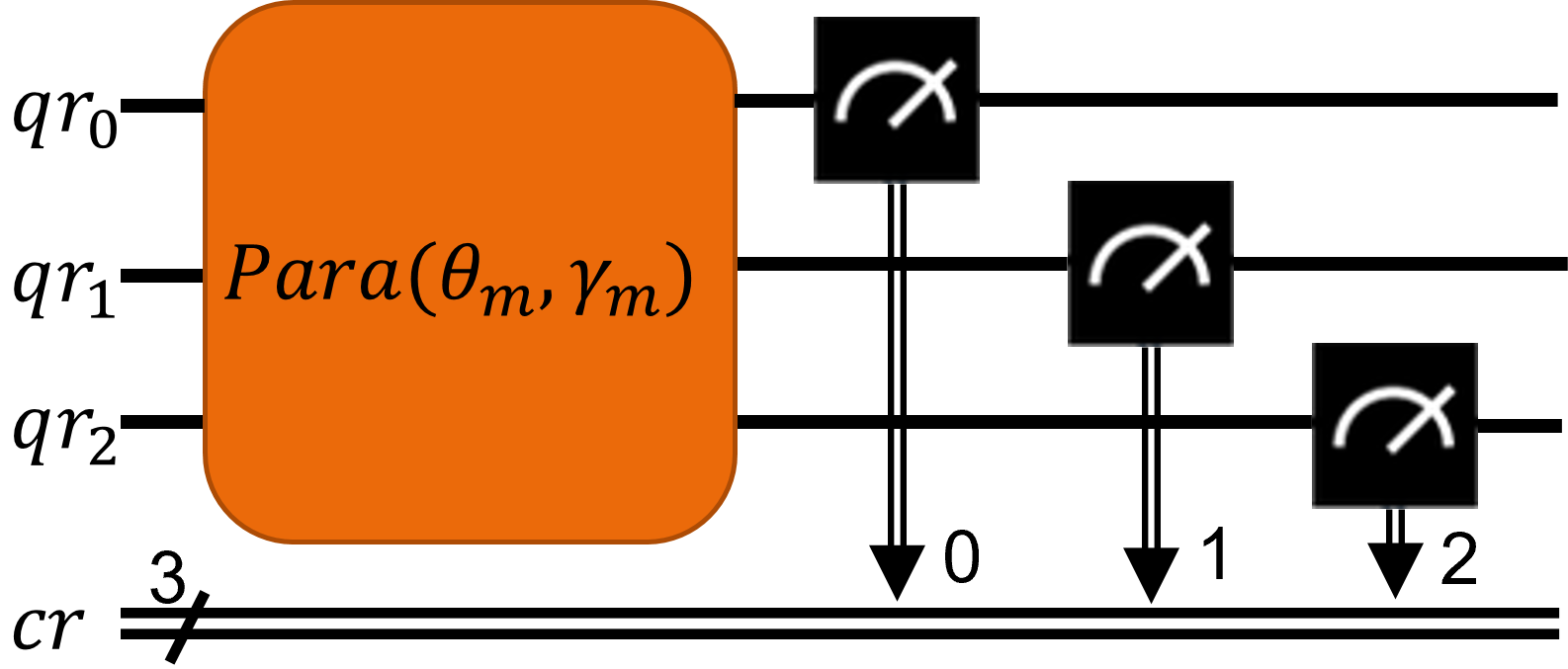}
  \end{minipage}%
  \begin{minipage}{.1\linewidth}
    \vspace{-0.5cm}
    \begin{eqnarray*}
       \hspace{0.5cm}\equiv
    \end{eqnarray*}
  \end{minipage}%
  \begin{minipage}{.38\linewidth}
    \includegraphics[width=0.99\textwidth]{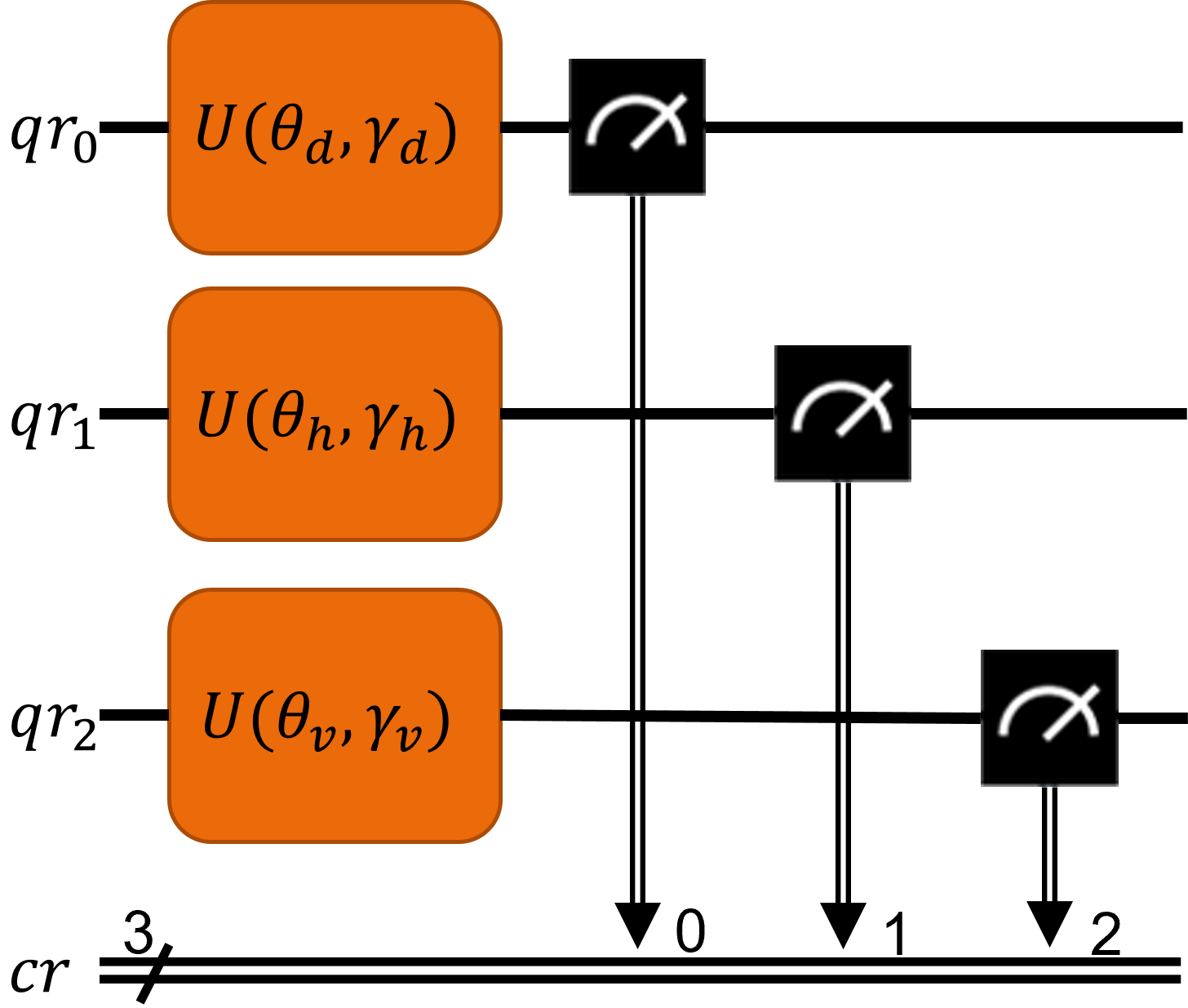}
  \end{minipage}
  \caption{Circuit scheme for \methodfour. Quantum circuit for a two pixel input image patch. The first qubit encodes the pixels in diagonal, the second those in horizontal, and the third those in vertical direction. We have $\gamma_m=(0,\pi)$, for all $m\in\{h,v,d\}$, since in all cases the first pixel of the filter mask is black and the second white (see Figures~\ref{fig:geng_alexander:one_dim_filter_horizontal} and \ref{fig:geng_alexander:one_dim_filter_vertical}). The \Hadamard- and \Phase-gates must be converted further into basis gates. In total, six \SX- and nine \Rz-gates are needed for this circuit.}
  \label{fig:geng_alexander:circuit_parallel}
\end{figure}

In the fourth variant, \methodfour, we combine the three directions in parallel instead of sequentially as before and marginalize the counts as in the third variant. Figure~\ref{fig:geng_alexander:circuit_parallel} yields the circuit for this variant.

\begin{figure}[tb]
	\centering
	\includegraphics[width=0.7\textwidth]{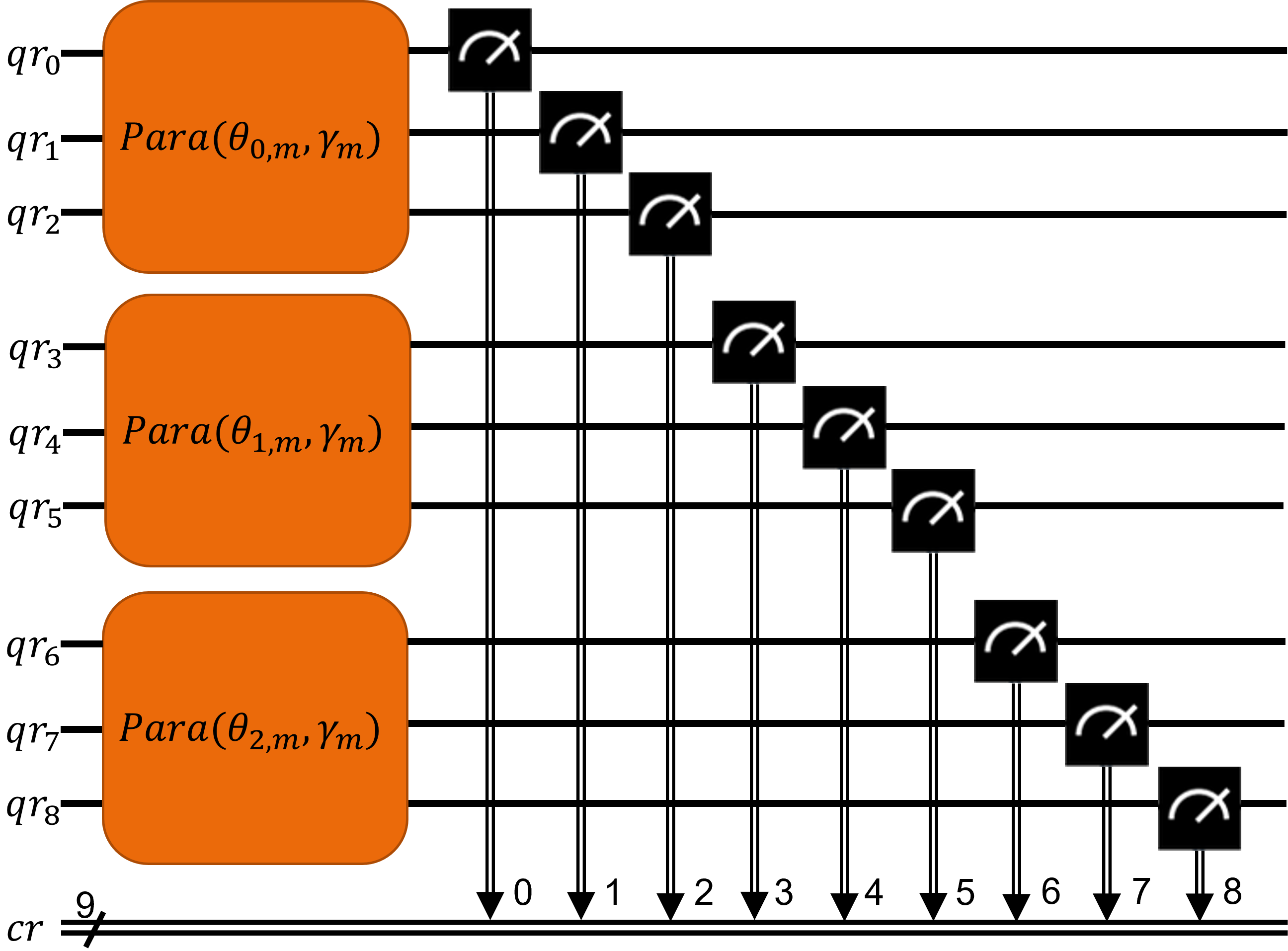}
	\caption{Circuit scheme for \methodfive. We apply \methodfour\, three times in parallel. The first index of the angle $\theta$ is the result $\{0,1,2\}$ of calculating the actual position of the input image patch modulo $3$. That means, the first input image patch is encoded in the qubits $qr_0-qr_2$, the second in $qr_3-qr_5$, the third in $qr_6-qr_8$, the fourth again in $qr_0-qr_2$, and so on. The second index describes the direction $m\in \{h,v,d\}$. It holds $\gamma_m=(0,\pi)$, for all $m\in\{h,v,d\}$, since in all cases the first pixel of the filter mask is black and the second white (see Figures~\ref{fig:geng_alexander:one_dim_filter_horizontal} and \ref{fig:geng_alexander:one_dim_filter_vertical}). In total, 18 \SX- and 27 \Rz-gates are needed for this circuit, if we decompose the Hadamard and Phase gates into basis gates.}
	\label{fig:geng_alexander:circuit_parallel3}
\end{figure}

Thanks to the parallel execution, we need less time to apply all gates than in \methodthree. However, we need three qubits instead of one. The fifth variant, \methodfive, extends \methodfour's main idea to more pixels. Instead of encoding only one pixel per circuit we parallelize the scheme for three pixels in one circuit as shown in Figure~\ref{fig:geng_alexander:circuit_parallel3}.

With this adaption, we triple the number of required qubits but simultaneously divide the number of required circuits by three. Clearly, this idea can be extended to more qubits, but we refrain from exemplifying this here.

\begin{figure}[tb]
	\centering
	\includegraphics[width=\textwidth]{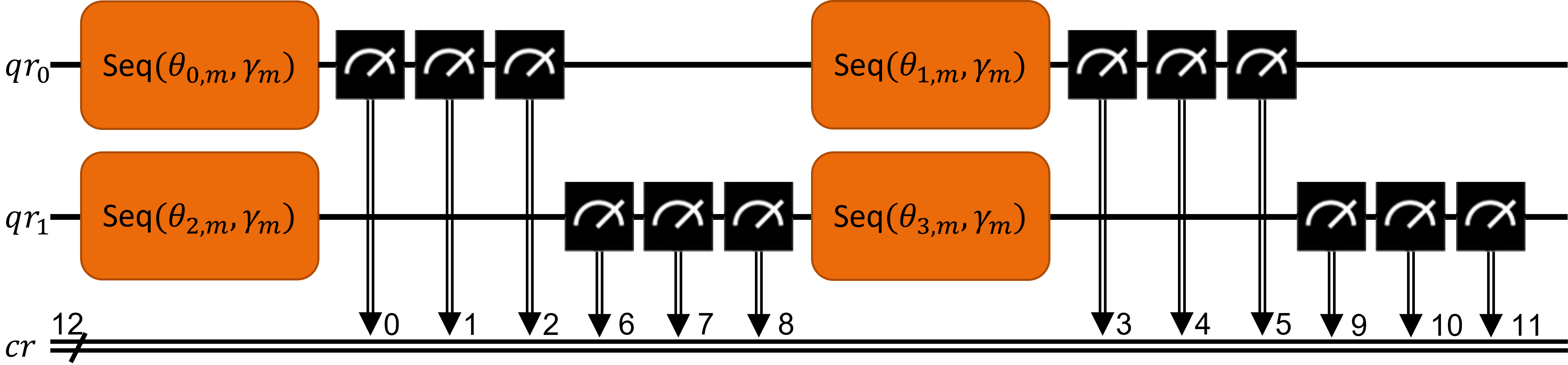}
	\caption{Circuit scheme for \methodsix. We apply \methodthree \,two times sequentially and two times in parallel. The first index of the angle $\theta$ is the actual position $\{0,1,2,3\}$ of the input image patch modulo $4$. That means, the first input image patch for all three directions is encoded in the top left Seq block, the second in the top right, the third in bottom left, and the forth in the bottom right block. The second index describes the direction $m\in \{h,v,d\}$. We have $\gamma_m=(0,\pi)$, for all $m\in\{h,v,d\}$, since in all cases the first pixel of the filter mask is black and the second white (see Figures~\ref{fig:geng_alexander:one_dim_filter_horizontal} and \ref{fig:geng_alexander:one_dim_filter_vertical}). In total, 24 \SX- and 36 \Rz-gates are needed for this circuit, if we decompose the Hadamard and Phase gates into basis gates.}
	\label{fig:geng_alexander:circuit_parallel_mid_2x2}
\end{figure}

Finally, a mixture of \methodthree\, and \methodfour\, leads to the sixth and last variant, that we cover in this paper, \methodsix. We take the mid-circuit measurement from \methodthree, but encode four pixel values in the three directions, instead of only one. That is,  we extend \methodthree\, by two pixels per qubit and parallelize this scheme on two qubits. The circuit for \methodsix\, is shown in Figure~\ref{fig:geng_alexander:circuit_parallel_mid_2x2}.

This way we divide the number of required circuits by 12 compared to \methodone \,and \methodtwo. We need 12 measurements per circuit for the four pixels. Clearly, the idea of this method can be extended for more pixels, both by more qubits and by more operations per qubit.

To compare the outcomes of the six variants, we use the Hellinger fidelity, which is defined in Equation~\eqref{equ:hellinger_equation}. 

As reference image, we calculate the pixel-wise maximum of the horizontal, vertical, and diagonal direction of the corresponding analytical descriptions \eqref{equ:analytical_result}. Note, that we use the state $\ket{0}$ there. The state $\ket{1}$ could also be used but would return the inverse image, with black edges and white background. For the three directions $m\in\{h,v,d\}$, we have the analytical description

\begin{equation}\label{equ:analytical_solution}
    \lvert\bra{0}U(\theta_m,\gamma_m)\ket{0}\rvert^2=\frac{1}{4}\lvert e^{i\lambda_{j+1,m}}+e^{i\lambda_{j,m}}\rvert^2.
\end{equation}
where $\lambda_{j,m}=\theta_{j,m}-\gamma_{j}$ and $\lambda_{j+1,m}=\theta_{j+1,m}-\gamma_{j+1}$. 

The pixel-wise maximum of the three resulting images is the reference image. The gray value frequencies of this image enter the Hellinger fidelity as entries $q_j$. The frequencies of the outcome of the real backends are plugged into \eqref{equ:hellinger_equation} as $p_j$.

\section{Near-term quantum computers setting}\label{sec:qc_environment}
Here, we describe our setting for evaluating our method from the previous section. It includes software, a classical computer, and quantum computers.

\begin{figure}[tb]
	\centering
	\begin{subfigure}[tb]{.57\linewidth}
        \includegraphics[width=0.99\textwidth]{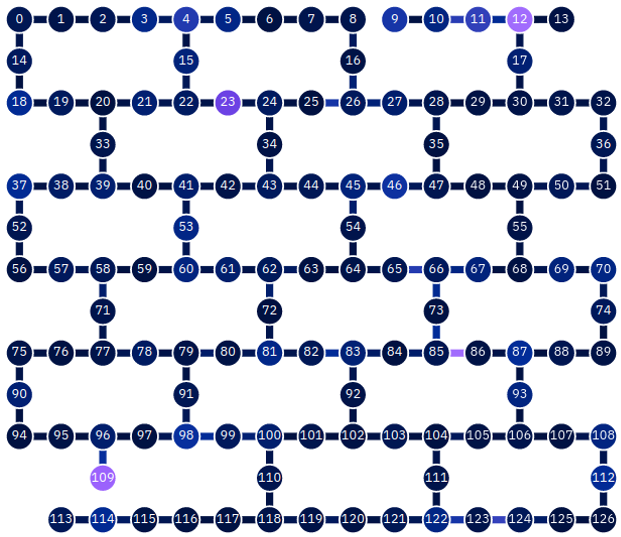}
        \caption{'ibm\_washington'}
    \end{subfigure}
    \begin{subfigure}[tb]{.42\linewidth}
        \centering
        \includegraphics[width=0.99\textwidth]{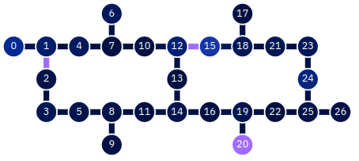}
        \caption{'ibm\_auckland'}
        \vspace{0.2cm}
        \centering
        \includegraphics[width=0.8\textwidth]{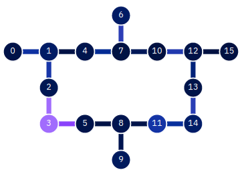}
        \caption{'ibmq\_guadalupe'}
    \end{subfigure}
	\caption{Coupling maps of backends used in this paper. Colors code the readout errors (points) and the \CX \,errors for the connections between the qubits (lines). Dark blue indicates a small error, purple a large one. We only show one example ('ibm\_auckland') for the backends with $27$ qubits (ibm\_auckland, 'ibmq\_mumbai', 'ibmq\_sydney', and 'ibmq\_ehningen') as the others look quite similar except for small deviations in the errors (see Table~\ref{tab:geng_alexander:error_rates}).}
	\label{fig:geng_alexander:used_backends}
\end{figure}

We use the open-source software development kit Qiskit \cite{qiskit_short} for working with IBM's circuit-based superconducting quantum computers \cite{ibm}. They provide a variety of systems, also known as backends, which differ in the type of the processor, the number of qubits, and their connectivity. Access is provided via a cloud. In this paper, we use the backends 'ibm\_auckland', 'ibm\_washington', 'ibmq\_guadalupe', 'ibmq\_mumbai', 'ibmq\_sydney', and 'ibmq\_ehningen'. The corresponding coupling maps are shown in Figure~\ref{fig:geng_alexander:used_backends}.

Additionally, the processor types and the performance values of the respective backend in terms of scale (number of qubits), quality (quantum volume), and speed (circuit layer operations per second [CLOPS]) are given in Table~\ref{tab:geng_alexander:actual_performance}.
\begin{table}[tb]
    \caption{Processor type and actual performance of the used backends as measured in December 2021. The backend 'ibmq\_ehningen' has currently no value for the speed.}
    \label{tab:geng_alexander:actual_performance}
    \centering
    \begin{tabular}{lclccc}
        \hline\noalign{\smallskip}
        Backend &&Processor& Scale & Quality & Speed\\
        &&type&[\# qubits] & [QV] & [CLOPS] \\
        \noalign{\smallskip}\hline\noalign{\smallskip}
        'ibm\_auckland'     &&Falcon r5.11& \phantom{1}27    & \phantom{1}64 & 2.400 \\
        'ibm\_washington'   &&Eagle r1& 127              & \phantom{1}32 & \hspace{0.22cm}850 \\
        'ibmq\_guadalupe'   &&Falcon r4P& \phantom{1}16    & \phantom{1}32 & 2.400 \\
        'ibmq\_mumbai'      &&Falcon r5.1& \phantom{1}27    & 128 & 2.000 \\
        'ibmq\_sydney'      &&Falcon r4& \phantom{1}27    & \phantom{1}32 & 1.800\\
        'ibmq\_ehningen'    &&Falcon r4& \phantom{1}27    & \phantom{1}32 & -\\
        \noalign{\smallskip}\hline
    \end{tabular}
\end{table}

Besides the various coupling maps and performance values, the backends underlie external influences. Characteristics of the backends, like CX error, readout error, or decoherence times, can change hourly. Calibration should diminish this effect, errors are however averaged over $24$ hours. Typical average values for CX error, readout error, decoherence times T1, T2, and frequency are shown in Table~\ref{tab:geng_alexander:error_rates}.
\begin{table}[tb]
    \caption{Typical average calibration data of the six chosen backends. The values are from December 2021.}
    \label{tab:geng_alexander:error_rates}
    \centering
    \begin{tabular}{lccccc}
        \hline\noalign{\smallskip}
        Backend & \CX \,error & Readout error & T1 & T2 & Frequency \\
        &[\%] & [\%] & [$\mu \rm s$] & [$\mu \rm s$] & [GHz]\\
        \noalign{\smallskip}\hline\noalign{\smallskip}
        'ibm\_auckland'                & 8.40 & 1.42 & 153.20    & 121.09  & 4.957 \\
        'ibm\_washington'   & 2.01 & 2.62 & \phantom{1}97.35    & \phantom{1}93.68  & 5.064 \\
        'ibmq\_guadalupe'        & 1.10 & 3.18 & 104.09              & \phantom{1}91.87            & 5.245 \\
        'ibmq\_mumbai'          & 4.63 & 2.46 & 132.78              & 124.18  & 4.880 \\
        'ibmq\_sydney'         & 1.29 & 9.12 & 104.82    & \phantom{1}94.73            & 4.960\\
        'ibmq\_ehningen'         & 1.05 & 2.09 & \phantom{1}99.99    & 122.32            & 5.164\\
        \noalign{\smallskip}\hline
    \end{tabular}
\end{table}

In addition to quantum computers, a classical computer is needed for preparing data and generating and storing the circuits before sending them to the quantum computer. We use a computer with an Intel Xeon E5-2670 processor running at $2.60$ GHz, a total RAM of $64$ GB, and Red Hat Enterprise Linux 7.9.

\section{Experimental results}\label{sec:application}
In this section, we show examples of what can be expected with current hardware for a classical edge detection task.

\subsection{Quantum edge detection with 2D mask}
\subsubsection{Binary image}
Starting with the experiment from Figure~\ref{fig:geng_alexander:scheme}, we use a $30\times 30$ binary sample image and two binary filter masks in horizontal and vertical direction (see Figures~\ref{fig:geng_alexander:two_dim_filter_horizontal} and \ref{fig:geng_alexander:two_dim_filter_vertical}). Black pixels are interpreted as an angle $0$ and white pixels as $\pi$. For each combination of input image patch and filter mask, we create one circuit. Thus, the $30\times 30$ sample image requires $900$ circuits for each direction. The results are interpretable and correct (see Figure~\ref{fig:geng_alexander:scheme} on the right side) even without error correction or mitigation techniques to reduce noise. 

\begin{figure}[tb]
	\centering
	\includegraphics[width=0.99\textwidth]{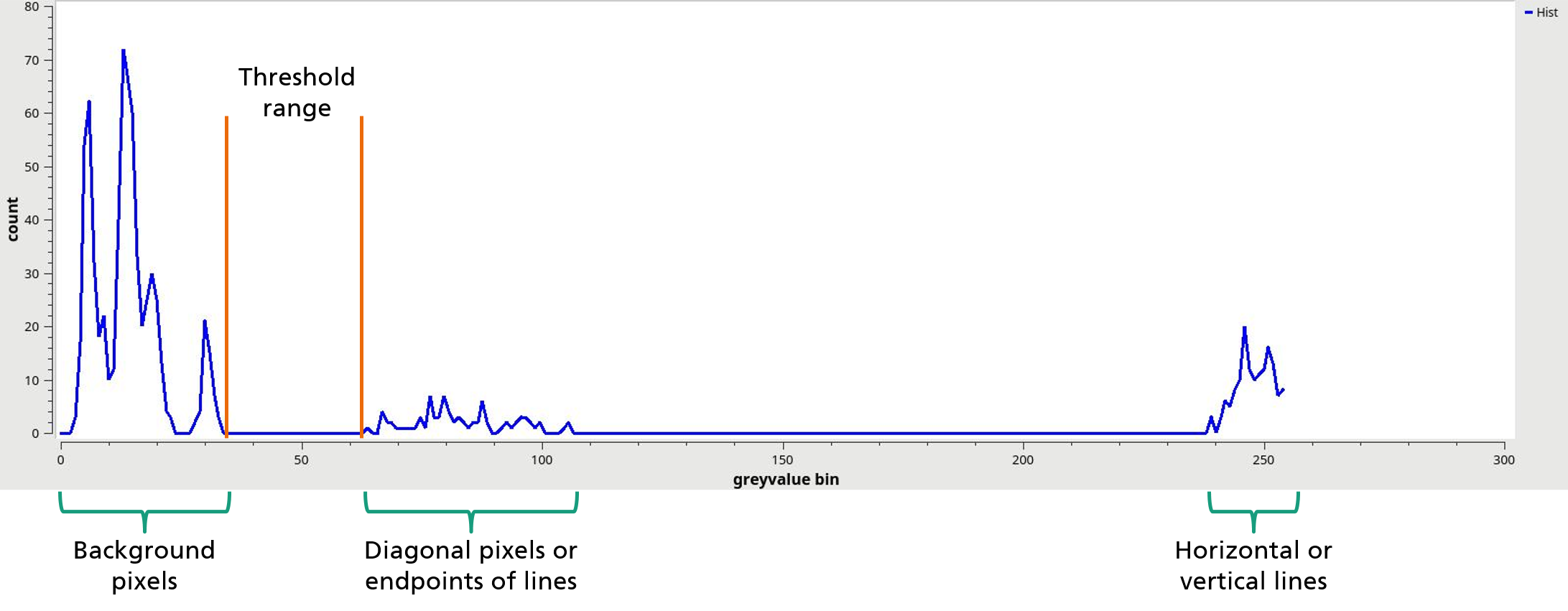}
	\caption{Typical histogram of the combined output image by applying the pixel-wise maximum of both directions. The threshold can be chosen in a threshold range even if the method is applied on a real backend (used backend here: ibmq\_ehningen).}
	\label{fig:geng_alexander:histogram}
\end{figure}

If we plot the histogram of the pixel-wise maximum of both directions $\voutb$ (see Figure~\ref{fig:geng_alexander:histogram}), the various types of pixels (edges, background, diagonals or endpoints of lines) are clearly distinguishable in three areas. Consequently, it is easy to choose a suitable threshold value. 

 For binary images, it is in theory also possible to use an approach based on the generation of hypergraph states similar to \cite{tacchino2019artificial} instead of the circuit given in Figure~\ref{fig:geng_alexander:circuit_2x2}. This is due to the fact that the prepared real equally weighted states like in Equation~\eqref{equ:geng_alexander_input_state} and \eqref{equ:geng_alexander_weight_state} ($\widetilde{\img_k}, \widetilde{\weight_k}\in \{-1,1\}$) coincide with the quantum hypergraph states \cite{rossi2013quantum}. By that, we can decrease the number of gates, especially the number of controlled gates. Since we deal with a quite small circuit, using the hypergraph states only yields a small improvement. For larger circuits, especially with multiple qubits that have to be entangled, the difference will be more pronounced.

\subsubsection{Gray value image}
\begin{figure}[tb]
	\centering
	\begin{subfigure}[tb]{.19\linewidth}
        \includegraphics[width=0.99\textwidth]{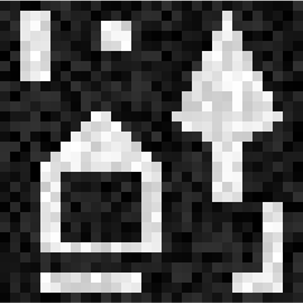}
        \caption{Sample}
        \label{fig:geng_alexander:gray_own_in}
    \end{subfigure}
	\begin{subfigure}[tb]{.19\linewidth}
        \includegraphics[width=0.99\textwidth]{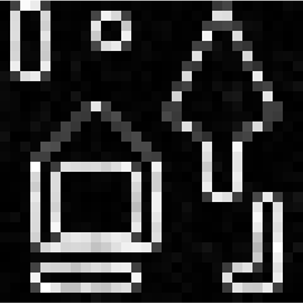}
        \caption{$\voutb$ sim}
        \label{fig:geng_alexander:gray_own_sim}
    \end{subfigure}
    \begin{subfigure}[tb]{.19\linewidth}
        \includegraphics[width=0.99\textwidth]{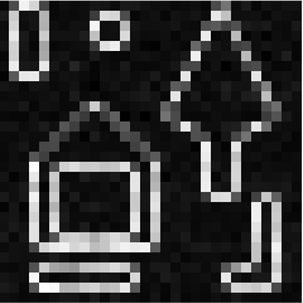}
        \caption{$\voutb$ back}
        \label{fig:geng_alexander:gray_own_real}
    \end{subfigure}
    \begin{subfigure}[tb]{.19\linewidth}
        \includegraphics[width=0.99\textwidth]{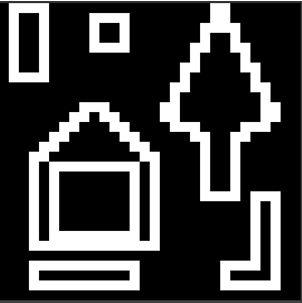}
        \caption{Edges sim}
        \label{fig:geng_alexander:gray_own_sim_edge}
    \end{subfigure}
    \begin{subfigure}[tb]{.19\linewidth}
        \includegraphics[width=0.99\textwidth]{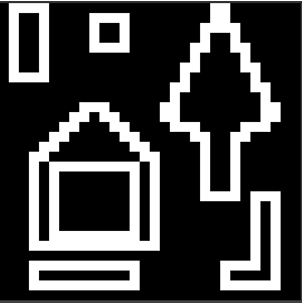}
        \caption{Edges back}
        \label{fig:geng_alexander:gray_own_real_edge}
    \end{subfigure}
	\caption{Results for a $30\times 30$ gray value image, created for test purposes. Only small deviation between the $\voutb$ of the 'qasm\_simulator' (sim) and backend 'ibmq\_ehningen' (back). All edges are detected in both cases.}
	\label{fig:geng_alexander:result_gray_own}
\end{figure}

As a toy example for a gray value image, we created a $30\times 30$ image (see Figure~\ref{fig:geng_alexander:gray_own_in}) with sharp edges. The quantum algorithm and the method are the same as above since the algorithm is already adapted to gray value images. We insert the angles (converted gray values as shown in Section~\ref{section:quantum_artificial_neuron}) into the quantum algorithm, get the results, and post-process as in the binary case. The outcomes are shown in Figure~\ref{fig:geng_alexander:result_gray_own}.

\begin{figure}[tb]
	\centering
	\includegraphics[width=0.99\textwidth]{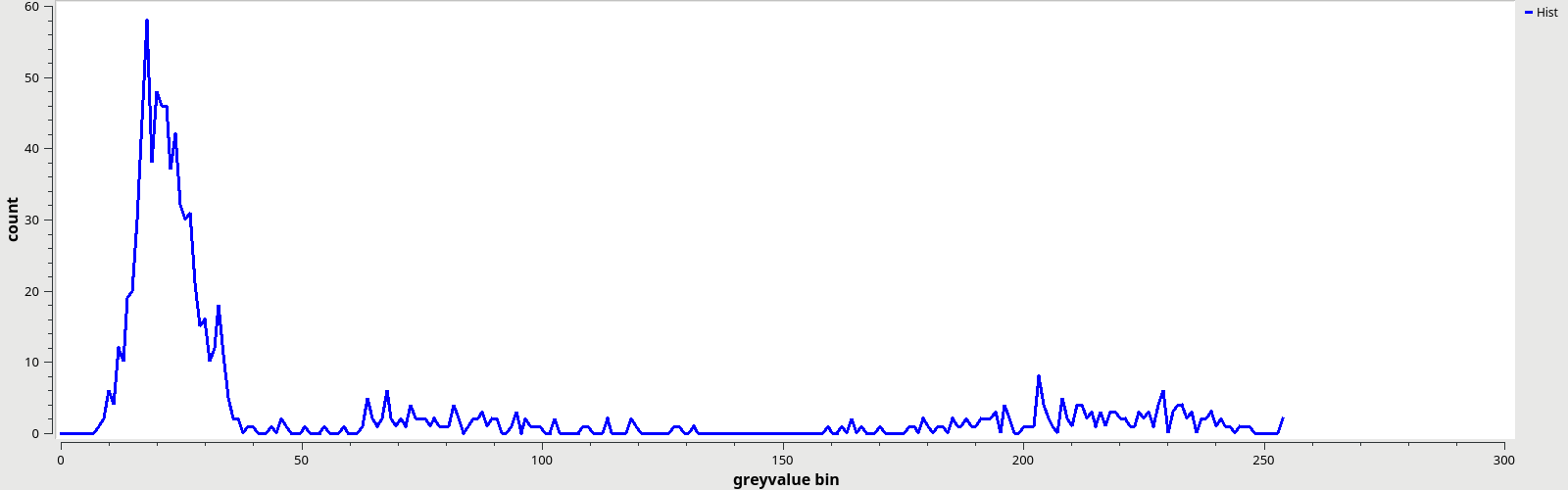}
	\caption{Typical histogram of the combined output image by applying the pixel-wise maximum of both directions (used backend here: ibmq\_ehningen).}
	\label{fig:geng_alexander:histogram_gray}
\end{figure}

Of course, the gray values affect the values of the outcome. Compared to Figure~\ref{fig:geng_alexander:scheme}, the $\voutb$ image in Figure~\ref{fig:geng_alexander:gray_own_sim} and \ref{fig:geng_alexander:gray_own_real} also shows lower values for the foreground and higher values for the background, which makes the threshold choice more difficult (see Figure~\ref{fig:geng_alexander:histogram_gray}). The three areas shown in Figure~\ref{fig:geng_alexander:histogram} are partly no longer distinguishable for all single pixels. However, it is still possible to detect all of the edges.

\begin{figure}[tb]
	\centering
	\begin{subfigure}[tb]{.19\linewidth}
        \includegraphics[width=0.99\textwidth]{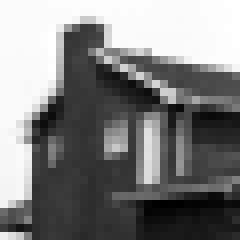}
        \caption{Sample}
        \label{fig:geng_alexander:house30_in}
    \end{subfigure}
	\begin{subfigure}[tb]{.19\linewidth}
        \includegraphics[width=0.99\textwidth]{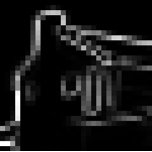}
        \caption{$\voutb$ sim}
        \label{fig:geng_alexander:house_30x30_sim}
    \end{subfigure}
    \begin{subfigure}[tb]{.19\linewidth}
        \includegraphics[width=0.99\textwidth]{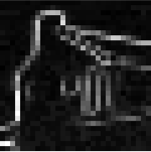}
        \caption{$\voutb$ back}
        \label{fig:geng_alexander:house_30x30_real}
    \end{subfigure}
    \begin{subfigure}[tb]{.19\linewidth}
        \includegraphics[width=0.99\textwidth]{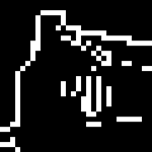}
        \caption{Edges sim}
        \label{fig:geng_alexander:house_30x30_sim_thres}
    \end{subfigure}
    \begin{subfigure}[tb]{.19\linewidth}
        \includegraphics[width=0.99\textwidth]{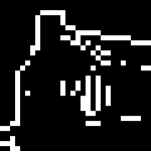}
        \caption{Edges back}
        \label{fig:geng_alexander:house_30x30_real_thres}
    \end{subfigure}
	\caption{Results for the downscaled $30\times 30$ House image of the USC-SIPI image database \cite{house}. Only small deviation between the $\voutb$ of the 'qasm\_simulator' (sim) and backend 'ibmq\_ehningen' (back).}
	\label{fig:geng_alexander:result_house30}
\end{figure}

Figure~\ref{fig:geng_alexander:result_house30} shows the outcomes for a downscaled classical image processing test image. The main edges in the image are detected.

\subsection{Quantum edge detection with 1D mask}\label{sec:results_one_dim}

\begin{figure}[tb]
	\centering
	\begin{subfigure}[tb]{.19\linewidth}
        \includegraphics[width=0.99\textwidth]{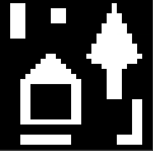}
        \caption{Sample}
        \label{fig:geng_alexander:binary_own_in}
    \end{subfigure}
	\begin{subfigure}[tb]{.19\linewidth}
        \includegraphics[width=0.99\textwidth]{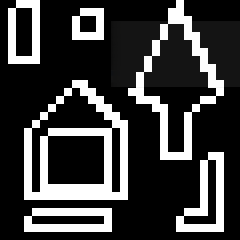}
        \caption{$\voutb$ sim}
        \label{fig:geng_alexander:binary_sim_one_dim}
    \end{subfigure}
    \begin{subfigure}[tb]{.19\linewidth}
        \includegraphics[width=0.99\textwidth]{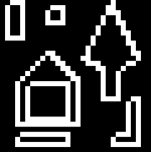}
        \caption{$\voutb$ back}
        \label{fig:geng_alexander:binary_real_one_dim}
    \end{subfigure}
    \begin{subfigure}[tb]{.19\linewidth}
        \includegraphics[width=0.99\textwidth]{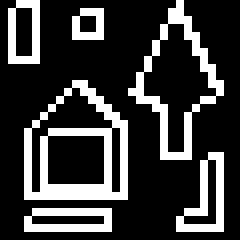}
        \caption{Edges sim}
        \label{fig:geng_alexander:binary_sim_one_dim_thres}
    \end{subfigure}
    \begin{subfigure}[tb]{.19\linewidth}
        \includegraphics[width=0.99\textwidth]{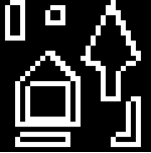}
        \caption{Edges back}
        \label{fig:geng_alexander:binary_real_one_dim_thres}
    \end{subfigure}
	\caption{Results for our $30\times 30$ binary test image with one-dimensional filtering. Only small deviation between the $\voutb$ of the 'qasm\_simulator' (sim) and backend 'ibmq\_ehningen' (back) with $32.000$ shots. Due to the one-dimensional filtering, pixels at the top left corner of the objects are not detected as edges.}
	\label{fig:geng_alexander:result_binary_own_one_dim}
\end{figure}

As in the two-dimensional case, we move the two-pixel sliding window through the whole image. For each step we create one circuit as visualized in Figure~\ref{fig:geng_alexander:circuit_1x2}. In total, we have the same amount of circuits needed to encode the image. That is, $900$ circuits for a $30\times 30$ gray value image. The outcome for the binary sample image (see Figure~\ref{fig:geng_alexander:scheme}) with the one-dimensional quantum edge detector is shown in Figure~\ref{fig:geng_alexander:result_binary_own_one_dim}. 
The method is well suited to detect vertical and horizontal edges in the image. However, some connections between the detected edges are missing like that in the top left corner of the objects.

\begin{figure}[tb]
	\centering
	\begin{subfigure}[tb]{.3\linewidth}
	    \centering
        \includegraphics[width=0.7\textwidth]{Fig14a.png}
        \caption{Sample image}
        \label{fig:geng_alexander:binary_own_in_three}
    \end{subfigure}
    \begin{subfigure}[tb]{.3\linewidth}
        \centering
        \includegraphics[width=0.7\textwidth]{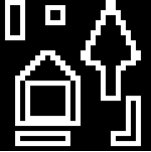}
        \caption{$\voutb$ backend \\}
        \label{fig:geng_alexander:binary_real_one_dim_three}
    \end{subfigure}
    \begin{subfigure}[tb]{.3\linewidth}
        \centering
        \includegraphics[width=0.7\textwidth]{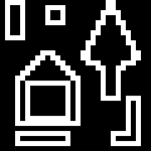}
        \caption{Edges backend}
        \label{fig:geng_alexander:binary_real_one_dim_thres_three}
    \end{subfigure}
	\caption{Results $\voutt$ for our $30\times 30$ binary test image with one-dimensional filtering in three directions. Solved the missing pixel problem of Figure~\ref{fig:geng_alexander:result_binary_own_one_dim}. Used backend 'ibmq\_ehningen'. Outcomes for 'qasm\_simulator' are omitted here, since there are no visual differences compared to the backend outcomes.}
	\label{fig:geng_alexander:result_binary_own_one_dim_three}
\end{figure}

This effect also holds for the diagonal edges of the house roof or the tree and explains the differences between the outcomes from Figure~\ref{fig:geng_alexander:scheme} and Figure~\ref{fig:geng_alexander:result_binary_own_one_dim}. With the adaption of Equation~\eqref{equ:max_total}, the missing edge pixels in Figure~\ref{fig:geng_alexander:result_binary_own_one_dim} are detected as shown in Figure~\ref{fig:geng_alexander:result_binary_own_one_dim_three}.

\subsection{Comparison of quantum edge detection with 1D and 2D mask}\label{sec:alexander_geng_comparison}
The main difference between the two variants is the size and depth of the quantum circuits. In the one-dimensional case, only one qubit and five gates are needed. In the two-dimensional case, we need three qubits, more gates, and especially the error-prone \CX-gates. If there is no connection between required qubits, additional SWAP-gates (three \CX-gates per SWAP-gate) are inserted in the transpilation step. Therefore the depth of the circuit on the real backend becomes larger. 

Since in the quantum edge detection with 1D filters we need fewer gates and no \CX-gates, it is also more robust to noise than that with 2D filters. The various combinations, which can occur, are a further reason. We calculate the inner product of the encoded input and weight quantum states. In the one-dimensional case, these are the two angles from the input image patch.

\begin{figure}[tb]
    \centering
    \resizebox{\linewidth}{!}{%
    \begin{tabular}{c|cc|cc}
    \large Sample images& \multicolumn{2}{c|}{\large One-dimensional}& \multicolumn{2}{c}{\large Two-dimensional}\\
    \includegraphics[height=3cm]{Fig14a.png} &  
    \includegraphics[height=3cm]{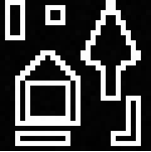}&
    \includegraphics[height=3cm]{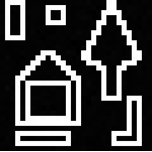} &
    \includegraphics[height=3cm]{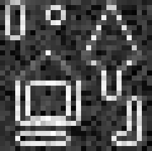} & \includegraphics[height=3cm]{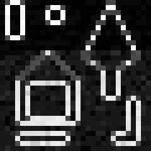}\\
     \includegraphics[height=3cm]{Fig11a.png} &  
    \includegraphics[height=3cm]{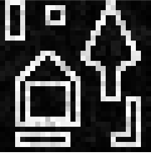}&
    \includegraphics[height=3cm]{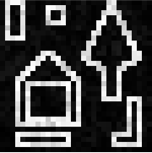} &
    \includegraphics[height=3cm]{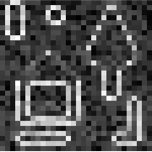} & \includegraphics[height=3cm]{Fig11c.png}\\
     \includegraphics[height=3cm]{Fig13a.png} &  
    \includegraphics[height=3cm]{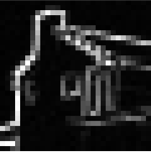}&
    \includegraphics[height=3cm]{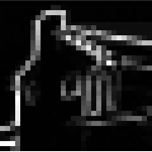} &
    \includegraphics[height=3cm]{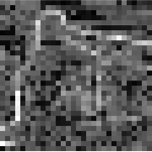} & \includegraphics[height=3cm]{Fig13c.png}\\
    & \large $1.000$ shots& \large $32.000$ shots&\large $1.000$ shots&\large $32.000$ shots\\

    \end{tabular}}
    \caption{Comparison one-dimensional and two-dimensional quantum edge detector with three $30\times 30$ sample images and $1.000$ or $32.000$ shots. Used backend 'ibmq\_ehningen'.}
	\label{fig:geng_alexander:comparison1qubit3qubit}
\end{figure}

In the two-dimensional case, we have three angles for the input image patch and three angles for the weights. By that, we have more classes (see, for example, Figure~\ref{fig:geng_alexander:histogram}). Not all of them can be distinguished from each other with a simple threshold value. Especially for gray value images, the values for edges can be indistinguishable from those of the background with noise. This effect is visualized in Figure~\ref{fig:geng_alexander:comparison1qubit3qubit}, especially with a lower number of shots ($1.000$ shot).

For the one-dimensional quantum edge detector, there are only small visual differences between the results with $1.000$ shots and those with the maximum number of $32.000$ using for example the backend 'ibmq\_ehningen'. The edges are visible and not strongly influenced by noise. This is not true for the two-dimensional variant. The more gates and the resulting errors make edge detection difficult, especially for the $30\times 30$ House image (see the bottom row in Figure~\ref{fig:geng_alexander:comparison1qubit3qubit}). The number of shots there is not sufficient to handle these errors. With a higher number of shots, like the $32.000$, it is possible.

The execution time depends linearly on the number of shots. Thus, reducing the number of shots is a good way to reduce execution times. For example, the quantum edge detection with $32.000$ shots nearly takes $43$ minutes per job (assuming that $300$ circuits can be processed per job), where $1.000$ shots only require $90$ seconds when using the backend 'ibmq\_ehningen'. Consequently, with the one-dimensional quantum edge detector, more jobs can be executed in the same time interval with usually better results as shown for example in Figure~\ref{fig:geng_alexander:comparison1qubit3qubit}.

\subsection{Improved quantum edge detection with 1D mask}\label{sec:alexander_geng_further_improvements}
Table~\ref{tab:geng_alexander:variants_one_dimensional} summarizes the six variants of the one-dimensional quantum edge detector. For the comparison, we take a $30\times 30$ gray value image as reference and assume that $300$ circuits can be executed per job on the real backends. This was the case for IBM's advanced backends in December 2021.

\begin{table}[tb]
    \caption{Summary of the six variants for the one-dimensional quantum edge detector at the example of a $30 \times 30$ gray value image. We take $300$ circuits per job as reference number (limit of IBM's advanced backends November 2021). To determine the execution times, we repeated the calculations on IBM's backend 'ibm\_auckland' 24 times between 22 and 29 December 2021.}
    \label{tab:geng_alexander:variants_one_dimensional}
    \centering
    \begin{tabular}{lrcrcr}
        \hline\noalign{\smallskip}
        Variant & shots & qubits & circuits & jobs & execution \\
         &  &  &  &  & time [s]\\        \noalign{\smallskip}\hline\noalign{\smallskip}
        \methodone       & $32.000$ & $1$ & $2.700$ & $9$    &$22.804\pm 5,9$\\
        \methodtwo        & $50$ & $1$ & $ 2.700$ & $9$       &$101\pm 1,9$\\
        \methodthree     & $50$ & $1$ & $900$ & $3$          &$38\pm 0,9$\\
        \methodfour        & $50$ & $3$ & $900$ & $3$          &$36\pm 0,7$\\
        \methodfive        & $50$ & $9$ & $300$& $1$           &$14\pm 0,4$\\
        \methodsix & $50$ & $2$ & $225$ & $1$     &$17\pm 0,9$\\
        \noalign{\smallskip}\hline
    \end{tabular}
\end{table}
The six methods differ in the number of shots, in the number of qubits, the number of circuits, and therefore also in the number of jobs, which have to be submitted to IBM. 
As a consequence, the execution time on the real backends varies for the six variants, too. We see a slightly bigger reduction of the execution time when using only $50$ shots instead of $32.000$ due to the linear correlation of the number of shots and the execution time \cite{Qiskit-Textbook_short}. Using \methodthree\, or \methodfour, we decrease the number of jobs by a factor of three so also the execution time approximately decreases by that factor. Furthermore, the fewer circuits/jobs explain why \methodfive\, and \methodsix\, need even less time.

\begin{figure}[tb]
	\centering
	\includegraphics[width=\textwidth]{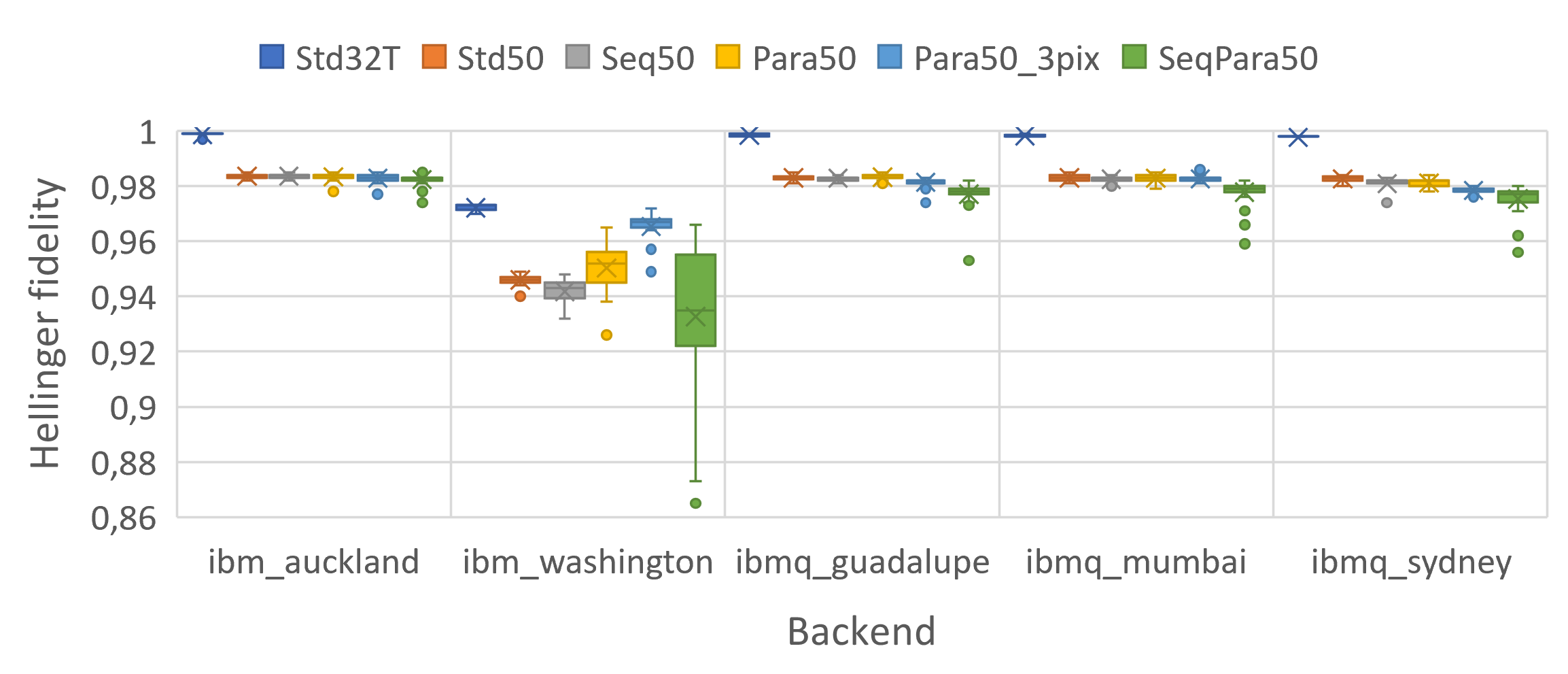}
	\caption{Hellinger fidelity for the six variants and five of IBM's backends. For the boxplot, we executed the codes three times a day between 22 and 29 December 2021 resulting in 24 runs.}
	\label{fig:geng_alexander:hellinger_fidelity}
\end{figure}

To compare the results of the six variant qualitatively, we use the Hellinger fidelity as defined in Section~\ref{sec:improved_1D}. Figure~\ref{fig:geng_alexander:hellinger_fidelity} contains boxplots of the fidelities for five backends.

The backends used in our study have more qubits than required for the six variants (see Figure~\ref{fig:geng_alexander:used_backends} and Table~\ref{tab:geng_alexander:variants_one_dimensional}). 
Since we do not apply any measurement error mitigation technique, we select the qubits with the lowest readout error.

All of the backends perform quite similarly, except for the newly released backend 'ibm\_washington'. One reason for this could be the release date right before the executions. For older backends, possible bugs have been discovered, whereas this may not yet be the case for 'ibm\_washington'. However, the improvement of the systems and especially of the newer systems is an ongoing process. Quantum computers get more stable and less error-prone with calibrations and adjustments. So, better results can be expected now already.

\begin{figure}[tb]
	\centering
	\begin{subfigure}[tb]{.32\linewidth}
        \centering
        \includegraphics[width=0.99\textwidth]{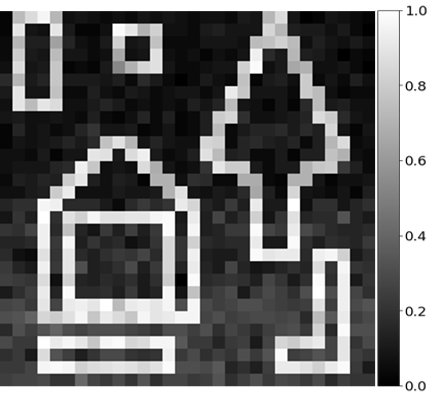}
        \caption{Worst case [0.86].}
        \label{fig:geng_alexander:output_method6a}
    \end{subfigure}
    \hspace{0.1cm}
	\begin{subfigure}[tb]{.32\linewidth}
	    \centering
        \includegraphics[width=0.99\textwidth]{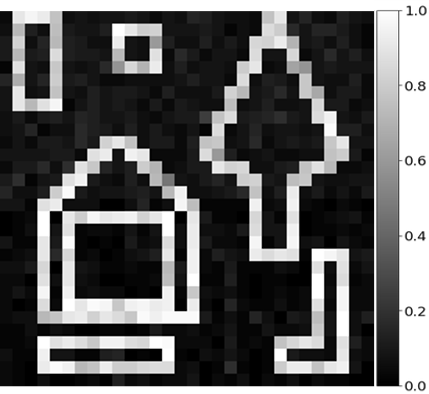}
        \caption{Best case [0.96].}
        \label{fig:geng_alexander:output_method6b}
    \end{subfigure}
    \hspace{0.1cm}
    \begin{subfigure}[tb]{0.28\linewidth}
        \centering
        \includegraphics[width=0.99\textwidth]{Fig15c.png}
        \caption{Detected edges.}
        \label{fig:geng_alexander:output_method6c}
    \end{subfigure}
	\caption{Worst and best results of 24 executions between 22 and 29 December 2021 for \methodsix\, using 'ibm\_washington' before applying a threshold. The value in brackets refers to the fidelity. Even with the worst result from all presented backends and variants of one-dimensional quantum edge detection, the edges are detectable by using an Otsu threshold afterwards as shown in \ref{fig:geng_alexander:output_method6c}.}
	\label{fig:geng_alexander:output_method6}
\end{figure}

Thanks to our method's robustness with respect to noise, even the low fidelity results from 'ibm\_washington' are completely interpretable. See Figure~\ref{fig:geng_alexander:output_method6} for the outcomes of \methodsix\, before and after applying a threshold. Some noise effects are visible for example in the background in Figure~\ref{fig:geng_alexander:output_method6a} with slightly higher gray values than expected. However, the foreground and background still differ sufficiently. Figure~\ref{fig:geng_alexander:output_method6c} shows the detected edges by using the worst case of \methodsix \, and 'ibm\_washington' backend, after applying an Otsu threshold. All edge pixels are detected even with the worst result of all experiments.

As expected, \methodone\, features the highest fidelity due to the high number of shots decreasing the effect of the finite-sampling shot noise. 

For the last five variants, we observe only slightly worse results and more variation in the results for \methodfive\, and \methodsix. A reason for that is the higher amount of measurements per circuit. We need $9$ and $12$ measurements per circuit for \methodfive\, and \methodsix, respectively. This increases the error per circuit and decreases the fidelity at the end.
\subsection{Larger images}\label{sec:larger_images}
\begin{figure}[tb]
	\centering
	\begin{subfigure}[tb]{.19\linewidth}
        \includegraphics[width=0.99\textwidth]{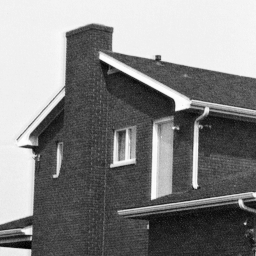}
        \caption{Sample}
        \label{fig:geng_alexander:house256_in}
    \end{subfigure}
	\begin{subfigure}[tb]{.19\linewidth}
        \includegraphics[width=0.99\textwidth]{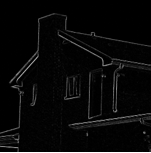}
        \caption{$\voutt$ sim}
        \label{fig:geng_alexander:house_256x256_sim}
    \end{subfigure}
    \begin{subfigure}[tb]{.19\linewidth}
        \includegraphics[width=0.99\textwidth]{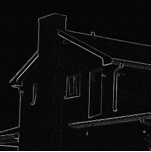}
        \caption{$\voutt$ back}
        \label{fig:geng_alexander:house_256x256_real}
    \end{subfigure}
    \begin{subfigure}[tb]{.19\linewidth}
        \includegraphics[width=0.99\textwidth]{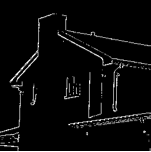}
        \caption{Edges sim}
        \label{fig:geng_alexander:house_256x256_sim_thres}
    \end{subfigure}
    \begin{subfigure}[tb]{.19\linewidth}
        \includegraphics[width=0.99\textwidth]{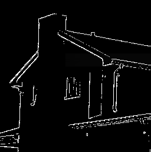}
        \caption{Edges back}
        \label{fig:geng_alexander:house_256x256_real_thres}
    \end{subfigure}
	\caption{Results for the $256\times 256$ House image \cite{house}. We use method \methodtwo, but all other five variants of the one-dimensional quantum edge detector yield similar results. Only small deviations of $\voutt$ from the 'qasm\_simulator' (sim) and backend 'ibm\_auckland' (back) are visible.}
	\label{fig:geng_alexander:result_house256}
\end{figure}

The advantage of our hybrid method is that for larger images the method itself and thus the basic errors remain the same. We create more circuits while keeping the size of the circuits the same. Therefore, our method is beneficial for practical usage in the current NISQ era. The House image with its original size of $256\times 256$ and the corresponding results are shown in Figure~\ref{fig:geng_alexander:result_house256}. Due to the findings from Section~\ref{sec:alexander_geng_comparison}, we exemplary use the one-dimensional variant \methodthree. 

The simulator and the backend outcomes differ only minimally, and the edges of the house are recognizable. This low noise error is mainly due to the very short quantum circuits. Thus, we can detect edges in arbitrarily large images with the current backends in today's NISQ era. 

Due to the limitations in the maximal number of circuits (exploratory and advanced $300$, core $900$, and open backends $100$ circuits per job at IBM as of November 8, 2021 \cite{ibm}), we split the circuits into several jobs and execute them sequentially. The jobs for the input image patches should be executed as directly consecutively as possible or at least with relatively equal calibrations. Otherwise, calibration variations show in the images, especially for larger quantum circuits with a lot of gates.
All six one-dimensional variants turned out to produce similar results. Hence, we only show the outcome of \methodtwo\, here.

Theoretically, it is also possible to process the entire $256\times256$ image in one circuit, e.g. with an extension of \methodsix. However, the number of measurements per job is currently limited. The exact number is not publicly available, but some own experiments have shown that about $16.000$ measurements per job are possible. For $2^a\times 2^a$ images, where $a\in \mathbb{N}$, this means a maximum image size of $64\times 64$ with \methodsix\, in one job. For larger images, we split the image into several parts and combine the results of multiple jobs classically afterwards.

\section{Conclusion and Discussion}\label{sec:conclusion}
In this paper, we practically implement a hybrid quantum edge detector in the current NISQ era. Starting from the quantum algorithm for an artificial neuron, we first develop a method that allows us to find edges in a gray value image using two-dimensional filter masks 
and replace these later by one-dimensional ones. This allows us to significantly reduce the circuit depth, the number of gates, and therefore also the influence of noise. Especially, we do not need any error-prone \CX \,gates. Due to this improvement, our method detects edges with a number of shots as low as $50$. This reduces execution time significantly.

We develop four additional variants of the one-dimensional quantum edge detection algorithm to adapt the method for larger images. In these, we consider several directions or pixels sequentially and/or in parallel, which leads to a reduction in the number of circuits. That way, we have to submit fewer jobs and can reduce the execution times further.

Of course, we are not limited to the presented variants. For example, we can encode in \methodthree\, more pixels sequentially or extend the idea of \methodfive\, further. Especially, \methodsix\, leaves space for customization. There, we use a $2\times 2$ pattern, where two pixels are encoded sequentially and repeat this for a second qubit. Instead of that, we can encode more pixels in one circuit. For example, we can implement a $16\times 16$ pattern. Thus, in each circuit $256$ pixels are encoded for all three directions. With that, we only need $256$ circuits to encode a $256\times 256$ image like the one in Figure~\ref{fig:geng_alexander:house256_in}. The number of circuits is in the range of allowed circuits per job. Thus, we theoretically need only one job on an IBM backend. 

Note that currently the total number of measurements per job is limited. This, for example, restricts the flexibility of \methodsix\, as not all pixels of a large image can be encoded in one job. Instead, the results of multiple jobs have to be combined classically afterwards.     

Each of the presented methods solves the quantum edge detection task. Other filtering tasks can be solved, too, by simply adapting the weights of the filter mask. For example, we can adapt the algorithm to enhance, denoise, or blur an image. 

To summarize, we implement a hybrid edge detector for larger images on a real quantum computer. To our knowledge, this has not been done before. The algorithmic idea based on quantum machine learning can be adapted flexibly to other tasks. This is a clear advantage compared to pure edge detection methods.

\newpage
\bibliographystyle{unsrt}
\bibliography{main}

\newpage
\section{Declarations}
\subsection{Funding}
This work was supported by the project AnQuC-3 of the Competence Center Quantum Computing Rhineland-Palatinate (Germany).
\subsection{Conflicts of interest/Competing interests}
The authors declare no competing interests.
\subsection{Availability of data and material}
All the data and simulations that support the findings are available from the corresponding author on request. 
\subsection{Code availability}
The jupyter notebooks used in this study are available from the corresponding author on request.
\subsection{Ethics approval}
\subsection{Consent to participate}
\subsection{Consent for publication}

\end{document}